\documentclass[Journal]{IEEEtran}

\usepackage{graphicx} 
\usepackage[dvips]{epsfig}  
\usepackage{mathptmx} 
\usepackage{times} 
\usepackage{amsmath} 
\usepackage{amssymb}  
\usepackage{amsmath}
\usepackage[dvips,bookmarksopen,bookmarksnumbered,citecolor=blue,urlcolor=blue]{hyperref}
\usepackage{psfrag}
\usepackage{enumerate,cite,latexsym,graphicx}
\newtheorem{theorem}{Theorem}[section]
\newtheorem{lemma}{Lemma}[section]

\newtheorem{definition}{Definition}[section]

\newtheorem{assumption}{Assumption}[section]
\newtheorem{claim}{Claim}
\newtheorem{rem}{Remark}[section]
\newtheorem{problem}{Problem}
\newcommand{\dfn}{\buildrel {\triangle}\over =}

\ifCLASSINFOpdf
\else
\fi

\begin{document}
%
\title{Lagrange Stabilization of  Pendulum-like Systems: A Pseudo $H_{\infty}$ Control Approach}
%
%
%
\author{Hua Ouyang,~\IEEEmembership{Member,~IEEE,}
        Ian R. Petersen,~\IEEEmembership{Fellow,~IEEE,}
        and~Valery ~Ugrinovskii,~\IEEEmembership{Senior~Member,~IEEE}
\thanks{The authors are with the University of New South Wales
at the Australian Defence Force Academy, Campbell, ACT 2600, Australia
e-mails: h.ouyang@adfa.edu.au, i.r.petersen@gmail.com, v.ugrinovskii@gmail.com.}
\thanks{Preliminary versions of the results of this paper were presented at the
  Joint 48th CDC and 28th CCC and the 2010 ACC.}
\thanks{This research is supported by the Australian Research Council.}}
\markboth{IEEE Transaction on Automatic Control, \textit{Manuscript for Review}}%
{Ouyang \MakeLowercase{\textit{et al.}}:  Manuscript for Review } \maketitle
\begin{abstract}
This paper studies the Lagrange stabilization of a class of nonlinear systems whose linear part has
a  singular system matrix and which have multiple periodic (in state) nonlinearities. Both state
and output feedback Lagrange stabilization problems are considered. The paper develops a pseudo
$H_{\infty}$
control theory to solve these stabilization problems. 
In a similar fashion to the Strict Bounded Real Lemma in classic $H_\infty$ control theory, a
Pseudo Strict Bounded Real Lemma is established for systems with a single unstable pole. Sufficient
conditions for the synthesis of state feedback and output feedback controllers are given to ensure
that the closed-loop system is pseudo strict bounded real. The pseudo-$H_\infty$ control approach
is applied to solve state feedback and output feedback Lagrange stabilization problems for
nonlinear systems with multiple nonlinearities. An example is given to illustrate the proposed
method.
\end{abstract}
\IEEEpeerreviewmaketitle
\section{Introduction}  \label{Sec:Introd}
The class of pendulum-like systems is a class of nonlinear systems with periodic (in state)
nonlinearities and an infinite number of  equilibria\cite{leonov1996fdm}. They cover an important
class of nonlinear systems arising in electronics, mechanics and power systems. These systems can
be used to model interconnected oscillators, synchronous electrical machines and electronic
phase-locked loop devices \cite{Shakh1982,Wang2006}. An important control objective in relation to
controlling such systems is to ensure that the closed-loop system retains the properties of a
pendulum-like system and its trajectories are bounded, at least, in the sense of Lagrange
stability. In combination with other analytical tools, this enables global asymptotic properties of
the system to be established. For example, the monograph \cite{leonov1996fdm} makes extensive use
of this approach to study global asymptotic behavior of nonlinear systems with periodic
nonlinearities and an infinite number of equilibria.

The concept of Lagrange stability can be traced back to H. Poincar\'{e}'s work in the 1890s
\cite{Poincare1957}. In \cite{Nemytskii1960}, Lagrange stability is defined as a property of a
state $x_0$ of a dynamical system $\dot{x}=f(t,x)$ given on a metric space $\mathcal{S}$, which
requires that the system trajectory $x=x(f,t,x_0)$ originating at this state $x_0$ to be contained
in a bounded set. It is shown in \cite{leonov1996fdm} that if a pendulum-like system 
possesses both
Lagrange stability and dichotomy, then it has a so-called gradient-like property. The gradient-like
property guarantees that any trajectory of the pendulum-like system eventually converges to an
equilibrium. This is analogous to the asymptotic stability of a system with a single equilibrium.
This observation highlights the importance of Lagrange stability  as a tool to establish the
gradient-like property of pendulum-like systems. It also motivates the study of pendulum-like
systems within the framework of Lagrange stability which is considered in this paper.

In the authors' previous work \cite{Ouyang08}, the state feedback controller synthesis problem is
considered for a restricted class of pendulum-like systems
in which the way that the controlled outputs  enter into the nonlinearities must have a special structure. 
In contrast to the results in \cite{Ouyang08}, this paper mainly focuses on solving the output
feedback Lagrange stabilization problem for pendulum-like systems with nonlinearities which have a general structure. 
Unlike the special case in \cite{Ouyang08}, in this more general case, a significantly different
method utilizing sign-indefinite solutions to game-type Riccati equations is necessary. This has
led us to develop a pseudo-$H_\infty$ control theory to address the Lagrange stabilization problem
of pendulum-like systems. This pseudo-$H_\infty$ control theory allows a pole of the closed-loop
transfer function to be located in the right half of the complex plane and ensures that the
closed-loop transfer function satisfies a frequency domain condition which is similar to the
bounded real property \cite{Anderson1973}. An important contribution of this paper is the pseudo
strict bounded real results in Theorems \ref{ESBRL0} and \ref{ESBRL00}, which are analogous to the
standard strict bounded real lemma \cite{Petersen1991}. Our pseudo-$H_\infty$ control theory can be
regarded as a theory which is analogous to the standard $H_\infty$ control theory (see
\cite{PUSB,Zhou1996}) but with a non-standard closed-loop stability condition. Furthermore, the
paper applies the proposed pseudo-$H_\infty$ theory to solve the Lagrange stabilization problem for
pendulum-like systems.

The usefulness of the Lagrange stability property of pendulum-like systems motivates research on
Lagrange stabilization of pendulum-like systems; e.g., see \cite{Wang2006,Yang2003,Li2005,Gao2009}.
 However, in these papers it was assumed that the nonlinear system contains a single nonlinearity only and has a
special matched structure on its nonlinearity. This special matched structure enables the Lagrange
stabilization problem to be cast as a standard $H_{\infty}$ problem. In order to consider general
system structures which do no satisfy matching conditions, a different approach is required which
motivates our pseudo $H_{\infty}$ control problem. Also, the results of
\cite{Wang2006,Yang2003,Li2005,Gao2009} are established using a Lagrange stability criterion given
 in \cite{leonov1996fdm} which requires  the linear part of the system to be minimal. This means that
 a post-check is required on the linear part of the resulting closed-loop system to determine if it is minimal.
  In contrast, this paper uses a Lagrange stability criterion which does not have the minimal realization
  requirement but uses a strict frequency-domain condition.  This Lagrange stability theory enables this paper to
consider a Lagrange stabilization problem without the requirement of a post-check on the minimality
of the linear part of the closed-loop system. Also, this Lagrange stability criterion allows us to
  solve the Lagrange stabilization problem for nonlinear systems with multiple nonlinearities. Indeed, a condition of the stability analysis techniques used in the paper is that the closed-loop
system matrix $A$ has a single zero eigenvalue, even though multiple nonlinearities are allowed.
The corresponding condition on the open-loop system in our control synthesis results is that this
system must have a single unobservable (or uncontrollable) mode at the origin.
%
%

To illustrate the efficacy of the proposed method, we give an example. It is concerned with
Lagrange stabilization of a network of three interconnected nonlinear pendulums. Also, this system
has some of the features of many practical systems such as power systems, large-scale
interconnected networks and hence it suggests some application areas for the theory developed in
this paper. These features are an interconnection of nonlinear but not identical elements, and the
existence of multiple equilibria points due to the periodicity of the nonlinear elements.

This paper is organized as follows: Section II formulates the Lagrange stabilization problem for
pendulum-like systems; Section III presents a pseudo $H_{\infty}$ control theory, which is
motivated by the problem formulated in Section 2; Section IV presents our main results on output
feedback Lagrange stabilization of unobservable pendulum-like systems; Section V presents our
 results on the output feedback Lagrange stabilization of uncontrollable pendulum-like systems;
Section VI gives results on the state feedback Lagrange stabilization of uncontrollable
pendulum-like systems. Section VII presents an example to illustrate the efficacy of the proposed
method and Section VIII concludes this paper. All of the proofs of the theorems in the Sections
II-VI are contained in the Appendix.

\emph{Notation:} $\mathcal{Z}$ denotes the set of integers. $\mathcal{R}^{n\times m}$ and
$\mathcal{C}^{n\times m}$ denote the space of ${n\times m}$ real matrices and the space of
${n\times m}$ complex matrices, respectively. $\mathcal{Q}$ denotes the set of rational numbers and
$\mathcal{Q}^m$ denotes the set of  vectors of $m$ rational numbers. $\sigma(A)$ denotes the set of
the eigenvalues of a matrix $A$.  $\sigma_{max}[\cdot]$ denotes the maximum singular value of a
matrix. $\mathcal{RH_{\infty}}$ denotes the space of all proper and real rational stable transfer
function matrices. $\mathcal{R}_+$ denotes the set of positive real numbers and
$\mathcal{R}^n_+=\left(\mathcal{R}_+\right)^n$. $\rho(X)$ denotes the spectral radius of the matrix
$X$. $\textrm{diag}[a_1,\cdots,a_n]$ is a diagonal matrix with $a_1,\cdots,a_n$ as its diagonal
elements. $\mathbb{B}(a,\epsilon)$ denotes a neighborhood around $a\in \mathcal{R}^n$, defined as
$\{\tilde{a}\in \mathcal{R}^n: \|\tilde{a}-a\|<\epsilon\}$. Given a vector
$\tau=\left[\tau_1,\cdots,\tau_m\right]^T\in\mathcal{R}^m_+$, $M_{\tau}$ denotes the diagonal
matrix $M_{\tau}=\textrm{diag}\left[\tau_1,\cdots,\tau_m\right]$. { Similarly,
$M_{\mu}=\textrm{diag}\left[\mu_1,\cdots,\mu_m\right]$.}   Given a vector $\nu\in\mathcal{Q}^m$,
$\rm{LCMD}(\nu)$ denotes the least common multiple (LCM) of the denominators of all the elements of
$\nu$.

\section{Problem Formulation of Lagrange Stabilization for Pendulum-like System}

%
\subsection{Pendulum-like Systems}
We consider  a class of nonlinear systems defined as follows:
\begin{eqnarray} \label{Ch6sys0}
   \dot{x} &=& Ax + B w, \nonumber \\
   z  &=& C x,
\end{eqnarray}
where $x\in\mathcal{R}^n$ is the state, $z\in\mathcal{R}^m$ is the nonlinearity output vector and $w\in\mathcal{R}^m$
is the nonlinearity input vector. Also, $A\in\mathcal{R}^{n\times n}$, $B\in\mathcal{R}^{n\times m}$,
$C=[C_{1}^T,\cdots,C_{m}^T]^T\in\mathcal{R}^{m\times n}$, $C_{i}\in\mathcal{R}^{1\times m},
i=1,\cdots, m$.  The components of the vector $w=[w_1,\cdots,w_m]^T$ are determined from the
corresponding components of the vector $z=[z_1,\cdots,z_m]^T$ via nonlinear functions
\begin{equation}\label{wi}w_i=\phi_i \left( {t,z_i} \right)\end{equation}  where $\phi_i:
\mathcal{R}_+\times\mathcal{R}\rightarrow\mathcal{R}$ is a  continuous, locally Lipschitz in the
second argument and periodic function with period $\Delta_i>0$; i.e.,
%
\begin{eqnarray}\label{nonlinearPeriod}
 \phi_i \left( {t, z_i  + \Delta_i } \right) = \phi_i \left(
{t,z_i } \right),  \quad \forall t\in \mathcal{R}_+, ~ z_i\in \mathcal{R}.
\end{eqnarray}
{ This type of nonlinearity appears frequently in the practical engineering systems mentioned in
Section \ref{Sec:Introd}. Phase-locked loops \cite{Leonov2006} and a pendulum system with a
vibrating point of suspension \cite{leonov1996fdm} are typical  examples of such
systems. We also refer to the example given in Section \ref{sec:Example}.} The transfer function of
the linear part of the system (\ref{Ch6sys0}) is given by $G(s)=C(sI-A)^{-1}B$. The nonlinear
functions $\phi_i\left(t, z_i \right), i=1,\cdots,m$, are assumed to satisfy the sector conditions,
\begin{equation}\label{Ch6sector_1}
-\mu_i   \le \frac{{\phi \left( {t, z_i} \right)}}{{z_i }} \le \mu_i,~ \forall t\in
\mathcal{R}_+,\quad z_i\neq 0,
\end{equation}
where $\mu_i \in \mathcal{R}_+, i = 1, \cdots, m$.

We define $\Delta\in \mathcal{R}^{m\times m}$ as $\Delta=\textrm{diag} [\Delta_1,\cdots,\Delta_m]$.
Given a vector $d \in\mathcal{R}^n$, let $\Pi(d)\dfn \{ k  d |k \in\mathcal{Z} \}$.
\begin{definition}\emph{(Pendulum-like System \cite{leonov1996fdm})} The nonlinear system (\ref{Ch6sys0}), (\ref{wi}),
(\ref{nonlinearPeriod}) is pendulum-like with respect to $\Pi(d)$ if for any solution $x(t, t_0,
x_0)$ of (\ref{Ch6sys0}), (\ref{wi}), (\ref{nonlinearPeriod}) with $x(t_0)=x_0$ , we
 have $x(t,t_0,x_0)+\bar{d}=x(t,t_0,x_0+\bar{d})$,
 for  all $t\ge t_0$, and all $\bar{d}\in \Pi(d)$.
\end{definition}
{ \begin{rem} This definition reflects the fact that the phase portrait of a pendulum-like system
is periodic. For example, in the case of a simple pendulum, this means that its position variable
can be represented by an angle between $0$ and $2\pi$.\end{rem} }
\begin{definition}\emph{(Lagrange Stability \cite{leonov1996fdm})}
The nonlinear system (\ref{Ch6sys0}), (\ref{wi}) is said to be Lagrange stable if all its solutions
are bounded.
\end{definition}
%
%
\subsection{Lagrange Stabilization Problem for Pendulum-like Systems}
\begin{figure}[htpb]
\begin{center}\psfrag{Ph}{$\phi_m$} \psfrag{Phi1}[c]{$~~\phi_1$}  \psfrag{Vdot}{$~~\vdots$}
\psfrag{u}{$u$} \psfrag{y}{$y$} \psfrag{W}{$w_1$} \psfrag{M}{$w_m$} \psfrag{Z1}{$z_1$}
\psfrag{N}{$z_m$}  \psfrag{K}[c]{$K(s)$}
\includegraphics[height=4cm]{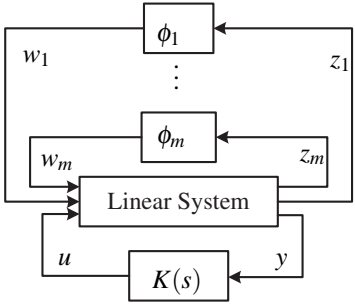}
 \caption{Nonlinear control system with periodic nonlinearities.} \label{Ch6FigO}
 \end{center}
 \end{figure}

 The pendulum-like system to be stabilized will be a controlled version of the nonlinear system (\ref{Ch6sys0}), (\ref{wi}), (\ref{nonlinearPeriod}), (\ref{Ch6sector_1}). That is, the linear part of the system is described by the state equations
\begin{subequations} \label{Ch6sys3} \begin{eqnarray}
\dot{x} & = & A x+B_2 u+B_1 w,\label{Ch6sys3a}\\
 z & =& C_1 x+D_{12}u,\label{Ch6sys3b}\\
 y &=& C_2 x+D_{21}w,\label{Ch6sys3c}
\end{eqnarray}\end{subequations} where $x\in \mathcal{R}^n$, $w\in\mathcal{R}^m$, $z\in\mathcal{R}^m$ are defined as in (\ref{Ch6sys0}),  $u\in\mathcal{R}^q$ is the
control input, and $y\in\mathcal{R}^p$ is the measured output.  Here, all the matrices are assumed
to have compatible dimensions. Also, the components of the nonlinearity input $w$ are related to
the components of the system output $z$ as in (\ref{wi}) and the nonlinearities $\phi_i$ have the
property (\ref{nonlinearPeriod}). Furthermore, the nonlinearities are assumed to satisfy the sector
condition (\ref{Ch6sector_1}).
The system block diagram is shown in Figure \ref{Ch6FigO}.

\begin{problem}\label{Problem1}\emph{(Output Feedback Lagrange Stabilization)}
  The output feedback Lagrange stabilization problem for the nonlinear system
(\ref{Ch6sys3}), (\ref{wi}), (\ref{nonlinearPeriod}), (\ref{Ch6sector_1}) is to design a linear
controller with the transfer function $K(s)$ and state-space realization:
 \begin{eqnarray}  \label{Ch6controller}
   \dot{x_c} &=& A_cx_c+B_cy \nonumber\\
   u &=& C_cx_c
 \end{eqnarray}
such that the resulting closed-loop system is pendulum-like and Lagrange stable.  
\end{problem}
\begin{problem}\emph{(State Feedback Lagrange Stabilization)} \label{Problem2} The state feedback Lagrange stabilization problem is to design a state feedback control law $u=Kx$ for
the system (\ref{Ch6sys3a}), (\ref{Ch6sys3b}), (\ref{wi}), (\ref{nonlinearPeriod}),
(\ref{Ch6sector_1}) to ensure that the resulting closed-loop system is pendulum-like and Lagrange
stable. 
\end{problem}

Note that in some cases, it may be possible to design a controller in the form of
(\ref{Ch6controller}) to asymptotically stabilize the system
 (\ref{Ch6sys3}), (\ref{wi}), (\ref{Ch6sector_1}). Such cases are trivial from the point of view of Lagrange stabilization.
 In order to rule out these trivial cases and to guarantee that the closed-loop system is a pendulum-like system, we will assume that the linear part of the systems (\ref{Ch6sys3}) has
 uncontrollable or unobservable modes.


To solve the above two problems, the following two technical results of \cite{Ouyang08} will be
used:
\begin{lemma}(\cite{Ouyang08}) \label{rationalPendulum}Consider the nonlinear system (\ref{Ch6sys0}), (\ref{wi}),
(\ref{nonlinearPeriod}). Suppose $\det A=0$ and there exists a vector $\bar{d}\neq 0$  such that
$A\bar{d}=0$, $C_{i}\bar{d}\neq 0, i=1,\cdots,m,$ and
$\left(\Delta\right)^{-1}C\bar{d}\in\mathcal{Q}^m$.
Also, let $\frac{\Delta_i}{C_{i}\bar{d}}=\frac{p_i}{q_i}$ for all $i=1,\cdots,m$, where $p_i$,
$q_i\neq 0$ are integers. Let $\bar{p}$ be the LCM of $p_i,~ i=1,\cdots,m$. Then, the system
(\ref{Ch6sys0}), (\ref{wi}), (\ref{nonlinearPeriod}) is pendulum-like with respect to $\Pi(d)$
where $d=\bar{p}\bar{d}$.\end{lemma}

\begin{lemma} (\cite{Ouyang08}) \emph{(Lagrange Stability Criterion)}  \label{LagrangeStability} Suppose the system (\ref{Ch6sys0}), (\ref{wi}),
(\ref{nonlinearPeriod}), (\ref{Ch6sector_1}) is a pendulum-like system. Also, suppose there exist a
constant $\lambda >0$ and a vector $\tau=\left[\tau_1,\cdots,\tau_m\right]^T\in\mathcal{R}^m_+$
satisfying the following conditions:
\begin{description}\item[i.] $A+\lambda I$ has $n-1$ eigenvalues with negative real parts and one with positive real part;
\item[ii.]  
${G} ^T \left( {-j\omega-\lambda } \right)M_{\tau}{G} \left( {j\omega-\lambda } \right) <
M_{\mu}^{-1}M_{\tau}M_{\mu}^{-1}$,
for all $\omega\ge 0$.
\end{description}
Then, the nonlinear system (\ref{Ch6sys0}), (\ref{wi}), (\ref{nonlinearPeriod}),
(\ref{Ch6sector_1}) is Lagrange stable.
\end{lemma}

The proofs of these two results appear in the journal version of \cite{Ouyang08} but are included
in the Appendix for completeness.


 Lemma \ref{LagrangeStability} is the key result to establish Lagrange stability of the closed-loop systems under consideration. It involves a
frequency domain condition, which is similar to the bounded real property in \cite{Anderson1973},
and a system state matrix $A+\lambda I$ which has one unstable eigenvalue. However, it does not
require the minimality of the linear part of the system (\ref{Ch6sys0}).  To establish these
conditions in the Lagrange stabilization problems 1 and 2, we develop a pseudo-$H_\infty$ control
theory in the next section, which is analogous to the standard $H_\infty$ control theory.
 \section{Pseudo-$H_\infty$ Control} \label{sec:PseudoHinfinity}
\subsection{The Pseudo Strict Bounded Real Property and the Corresponding Strict Bounded Real Lemma (SBRL)}

The bounded real property is an important concept frequently used in the standard $H_{\infty}$
control theory.  We begin our development of pseudo $H_{\infty}$ control  with the definition of
the pseudo strict bounded real property, which is analogous to the  standard bounded real
property.
\begin{definition} A matrix $A\in\mathcal{R}^{n\times n}$ which has $n-1$
eigenvalues with negative real parts and one eigenvalue with positive real part is said to be
\emph{pseudo-Hurwitz}.
A symmetric matrix $P\in \mathcal{R}^{n\times n}$ is said to be \emph{pseudo-positive definite} if
it has $n-1$ positive eigenvalues and one negative eigenvalue.
\end{definition}
 \begin{definition} A linear time-invariant (LTI) system (\ref{Ch6sys0}) is called \emph{pseudo strict
bounded real} if the following conditions hold: \begin{description}\item[(i)] $A$ is pseudo
Hurwitz;
\item[(ii)] \begin{equation}\label{VO.ppr} \max\limits_{\omega\in\mathcal{R}}\{ \sigma_{max}
[G(-j\omega)^T G(j\omega)]\}<1.\end{equation}
\end{description}
\end{definition}

\begin{theorem}
\label{ESBRL0} Consider the LTI system (\ref{Ch6sys0}). If the Riccati equation
  \begin{equation}\label{ARE0}
  A^TP+PA+PBB^TP+C^TC=0
  \end{equation} has a solution $P=P^T$ such that $P$ is pseudo-positive definite and
$A+BB^TP$ has no purely imaginary eigenvalues, then the system (\ref{Ch6sys0}) is pseudo strict
bounded real.
\end{theorem}

\begin{theorem}
\label{ESBRL00} If the LTI system (\ref{Ch6sys0}) is pseudo strict bounded real,  then
\begin{enumerate}
\item There exists a pseudo-positive definite matrix $P=P^T$ such that
\begin{equation}\label{Riccati_inequality}
A^TP+PA+PBB^TP+C^TC<0.
\end{equation}

\item Furthermore, if in addition the pair $(A,B)$ is stabilizable and the pair $(A,C)$
  is observable,  then the Riccati equation (\ref{ARE0})
has a stabilizing solution $P$ which is pseudo-positive definite.\end{enumerate}  \end{theorem}


Theorem~\ref{ESBRL0}  is analogous to the sufficiency part of the strict bounded real lemma for
systems with non-minimal realizations\cite{Petersen1991}. Also, Theorem \ref{ESBRL00} 
is analogous to the necessity part of the strict bounded real lemma for systems with non-minimal
realizations.
Theorems \ref{ESBRL0} and \ref{ESBRL00}  are together called the \emph{pseudo strict  bounded real
lemma}.

The pseudo strict bounded real lemma gives a relationship between state-space conditions, such as
solvability of (\ref{ARE0}) and pseudo-Hurwitzness of $A$,  and the frequency-domain inequality
(\ref{VO.ppr}). This will allow us to replace the frequency domain condition for the closed-loop
system that will appear in the application of Lemma \ref{LagrangeStability}, with a condition in the state-space form. This is a key step in the derivation of a solution to Problems 1 and 2.  
\subsection{State Feedback Pseudo-$H_\infty$ Control}


 The state feedback pseudo-$H_\infty$ control problem for the LTI system (\ref{Ch6sys3a}), (\ref{Ch6sys3b})
 involves designing a state feedback law $u=Kx$ which ensures that the corresponding closed-loop system
 is pseudo strict bounded real.
In an analogous way to $H_\infty$ control theory \cite{PUSB,Zhou1996},
 the main result of this section presented in the following theorem, gives  a
sufficient condition for the existence of a solution to the problem.

The following assumption is made on the system (\ref{Ch6sys3a}), (\ref{Ch6sys3b}):
\begin{assumption}\label{Assumption.A1}
$E_1=D_{12}^T D_{12}>0$.
\end{assumption}

\begin{theorem}\label{synthesis0} Suppose Assumption~\ref{Assumption.A1} holds for the system (\ref{Ch6sys3a}), (\ref{Ch6sys3b}) and the
  Riccati
equation
\begin{eqnarray}\label{H_controller}
&&(A
-B_2E_1^{-1}D_{12}^TC_1)^TP+P(A-B_2E_1^{-1}D_{12}^TC_1)\nonumber \\
&&+P(B_1B_1^T-B_2E_1^{-1}B_2^T)P+C_1^T(I-D_{12}E_1^{-1}D_{12}^T)C_1=0\nonumber \\
\end{eqnarray} has a solution $P=P^T$ such that $P$ is pseudo-positive definite and the matrix
\begin{equation}\label{stabilizingmatrix}A-B_2E_1^{-1}D_{12}^TC_1+(B_1
B_1^T-B_2E_1^{-1}B_2^T)P\end{equation} has no purely imaginary eigenvalues.
Then, the state feedback control law
\begin{equation}\label{controllaw}u=-E_1^{-1}(B_2^TP+D_{12}^TC_1)x\end{equation} solves the state feedback pseudo-$H_\infty$ control
problem. That is, the resulting closed-loop system
 is pseudo strict bounded real.

\end{theorem}
 \begin{rem}In practice, it is usually convenient to use the stabilizing solution to the
Riccati equation
(\ref{H_controller}) 
in order to construct the required state feedback control law (\ref{controllaw}).
\end{rem}
\subsection{Output Feedback Pseudo-$H_\infty$ Control}

 Analogous to the standard output feedback $H_\infty$ control problem,
 the output feedback
 pseudo-$H_\infty$ control
 problem for the system (\ref{Ch6sys3}) involves designing a compensator of the form (\ref{Ch6controller}) 
to make the corresponding closed-loop system pseudo strict bounded real. 
The following two theorems each give a
  sufficient condition for the existence of a solution to the output feedback
pseudo-$H_\infty$ control problem for a system of the form (\ref{Ch6sys3}). Besides Assumption
\ref{Assumption.A1}, the following assumption is also made on the system (\ref{Ch6sys3}):

\begin{assumption}\label{Assump1}
$E_2=D_{21}D_{21}^T>0$.
\end{assumption}

\begin{theorem}\label{theorem20}
Suppose the system (\ref{Ch6sys3}) satisfies Assumptions \ref{Assumption.A1} and \ref{Assump1} and
the following conditions are satisfied: \noindent\begin{description}
\item[(i)]\noindent The Riccati equation
\begin{eqnarray}
\label{AREp1}
&&(A-B_2 E_1^{-1} D_{12}^T C_1)^T X + X (A-B_2 E_1^{-1} D_{12}^TC_1) \nonumber\\
&&+ X(B_1B_1^T-B_2E_1^{-1}B_2^T)X \nonumber \\
&& + C_1^T(I-D_{12}E_1^{-1}D_{12}^T)C_1=0
\end{eqnarray}
has a stabilizing solution $X=X^T$ which is pseudo-positive definite;
\item[(ii)]\noindent The Riccati equation
\begin{eqnarray}
\label{AREp2}
&& (A-B_1D_{21}^TE_2^{-1}C_2)Y + Y(A-B_1D_{21}^TE_2^{-1}C_2)^T\nonumber \\
&&+Y(C_1^TC_1-C_2^TE_2^{
-1} C_2)Y\nonumber\\
&&+ B_1 (I-D_{21}^TE_2^{-1}D_{21})B^T_1=0
\end{eqnarray}
has a  stabilizing solution $Y=Y^T$ which is positive definite;
\item[(iii)] The matrix $XY$ has a spectral radius strictly less than one, $\rho(XY)<1$.
\end{description}

Then, there exists a dynamic output feedback compensator of the form (\ref{Ch6controller}) such
that the resulting closed-loop system
is pseudo strict bounded real. Furthermore, the matrices defining the required dynamic feedback
controller (\ref{Ch6controller}) can be constructed as follows:
\begin{eqnarray}\label{conParameters}A_c &=& A+B_2C_c-B_cC_2+(B_1-B_cD_{21})B_1^TX,\nonumber\\
B_c &=& (I-YX)^{-1}(YC_2^T+B_1D_{21}^T)E_2^{-1},\nonumber\\
C_c &=& -E_1^{-1}(B_2^TX+D_{12}^TC_1).\end{eqnarray}
 \end{theorem}
\begin{theorem}\label{theorem200}
Suppose the system (\ref{Ch6sys3}) satisfies Assumptions \ref{Assumption.A1} and \ref{Assump1} and
the following conditions are satisfied:
\begin{description}\item[(i)] The Riccati equation (\ref{AREp1}) has a positive
definite stabilizing solution $X=X^T$;
\item[(ii)]  the Riccati equation (\ref{AREp2}) has a pseudo-positive definite stabilizing solution
  $Y=Y^T$ ;
\item[(iii)] The matrix $XY$ has a spectral radius strictly less than one, $\rho(XY)<1$.
\end{description}
Then, there exists a dynamic output feedback compensator of the form (\ref{Ch6controller}) such
that the resulting closed-loop system
is pseudo strict bounded real. Furthermore, the matrices in the required dynamic feedback
controller (\ref{Ch6controller}) can be constructed as follows:
\begin{eqnarray} \label{controllerParameters}
A_c &=& A+B_cC_2-B_2 C_c + YC_1^T(C_1- D_{12}C_c),\nonumber \\
 B_c &=& -(YC_2^T+B_1^T D_{21}^T)E_2^{-1},\nonumber\\
C_c &=& E_1^{-1}(B_2^T X + D_{12}^TC_1)(I-Y X)^{-1}.\end{eqnarray}
\end{theorem}
{\begin{rem} According to \cite{Basar1995}, the stabilizing solutions to the Riccati equations
(\ref{AREp1}) and (\ref{AREp2}) are unique, if they exist.  \end{rem} }


 \section{Output Feedback Lagrange Stabilizing Controller Synthesis for Unobservable
 Systems} In this section, the output feedback pseudo $H_{\infty}$
control theory developed in the previous section is used to solve Problem \ref{Problem1} for
nonlinear systems satisfying the following assumptions, which will be used to ensure that the
closed-loop system is pendulum-like and to rule out trivial cases in which the nonlinear system can
be asymptotically stabilized:
\begin{assumption}\label{Ch6.Assum.Unobserv}  There
exists a non-zero vector $x$ such that $Ax=0$ and $C_2x=0$.
\end{assumption}
%
%
%

Assumption \ref{Ch6.Assum.Unobserv} implies that $(A, C_2)$ is unobservable and the origin is an
unobservable mode. Using the Kalman decomposition in the unobservable form \cite{Antsaklis2006}, it
follows that there exists a non-singular state-space transformation matrix $T$ such that the system
matrices of the system (\ref{Ch6sys3}) are transformed to the form
 \begin{eqnarray} \label{Ch6.KD.c}\tilde{A}  &=& T^{-1}AT=\left[
\begin{array}{cc}\tilde{A}_1 & 0\\\tilde{A}_{2} & 0\end{array}\right],~
\tilde{B}_2=T^{-1}B_2=\left[\begin{array}{c}\tilde{B}_{2a}\\\tilde{B}_{2b}\end{array}\right],
\nonumber \\
\tilde{B}_1 &=&  T^{-1} B_1 =  \left[\begin{array}{c}\tilde{B}_{1a} \\
 \tilde{B}_{1b}\end{array}\right],\nonumber\\
 \tilde{C}_1 &=& C_1T =  \left[\begin{array}{cc}\tilde{C}_{1a} & \tilde{C}_{1b}\end{array}\right],~
 \tilde{D}_{12}  =  D_{12}, \nonumber \\
 \tilde{C}_{2} &=&  C_{2} T = \left[\begin{array}{cc}\tilde{C}_{2a} & 0\end{array}\right],~\tilde{D}_{21}  =
 D_{21},
 \end{eqnarray} where $\tilde{A}_1\in\mathcal{R}^{(n-l)\times
(n-l)}$, $\tilde{B}_{1a}\in \mathcal{R}^{(n-l)\times m}$, $\tilde{B}_{2a}\in
\mathcal{R}^{(n-l)\times q}$,
$\tilde{C}_{1a},~\tilde{C}_{2a}\in\mathcal{R}^{m\times (n-l)}$. 

Also, let $e_n= [\begin{array}{cc} 0_{1\times (n-1) } & 1\end{array} ]^T$. We define two vectors
$\chi={C}_1Te_n $ and $\bar{d}=[0_{1\times n} ~ e_n^T T^T]^T\in\mathcal{R}^{2n}$.

\begin{assumption}\label{W.rational} There exists a constant $\tau_0>0$ such that all the elements of the vector $\nu=\tau_0\Delta^{-1}\chi$
are non-zero rational numbers.  \end{assumption}

\begin{rem} In the case where the coefficients in the system (\ref{Ch6sys3}) are all rational numbers,
Assumption \ref{W.rational} amounts to an assumption that the periods of the nonlinearities are commensurate. 
\end{rem}

 The main result of this section involves the following Riccati equations dependent on parameters $\lambda>0$ and $\tau_i>0,~i=1,\cdots, m$:
\begin{eqnarray} \label{Ch6ARE1}  \lefteqn{(\lambda I+A-B_2 \bar{E}_1^{-1} D_{12}^T M_{\tau} C_1)^T
X + X (\lambda I+A-B_2 \bar{E}_1^{-1} D_{12}^T M_{\tau}C_1)}\nonumber\\
&& +X(B_1M_{\mu}M_{\tau}^{-1}M_{\mu}B_1^T-B_2\bar{E}_1^{-1}B_2^T)X\nonumber \\
&&+
C_1^T(M_{\tau}-M_{\tau}D_{12}\bar{E}_1^{-1}D_{12}^T M_{\tau})C_1=0, \hspace{2.5cm}
\end{eqnarray}
\begin{eqnarray}
\label{Ch6ARE2} 
\lefteqn{(\lambda I+A-B_1M_{\mu}M_{\tau}^{-1}M_{\mu} {D}_{21}^T\bar{E}_2^{-1}C_2)Y}
\nonumber \\
&&+
Y(\lambda
I+A-B_1M_{\mu}M_{\tau}^{-1}M_{\mu} {D}_{21}^T\bar{E}_2^{-1}C_2)^T\nonumber\\
&&+B_1\left( \begin{array}{l}M_{\mu}M_{\tau}^{-1}M_{\mu}B^T_1\\
   -M_{\mu}M_{\tau}^{-1}M_{\mu}D_{21}^T\bar{E}_2^{-1}D_{21}M_{\mu}M_{\tau}^{-1}M_{\mu}\end{array}\right)B^T_1\nonumber\\
&&+Y(C_1^TM_{\tau}C_1-C_2^T\bar{E}_2^{-1} C_2)Y=0, 
\end{eqnarray}
where $\bar{E}_1=D_{12}^TM_{\tau}D_{12}$ and $\bar{E}_2=D_{21}M_{\mu}M_{\tau}^{-1}M_{\mu}D_{21}^T$.
If these Riccati equations have suitable solutions, we will define the parameter matrices of the
controller (\ref{Ch6controller})  as
follows: 
 \begin{eqnarray}\label{Ch6AREparameter.B}A_c &=& A+B_cC_2-B_2 C_c + YC_1^T(M_{\tau}C_1- M_{\tau}D_{12}C_c),\nonumber \\
 B_c &=& -( Y C_2^T+ B_1M_{\mu} M_{\tau}^{-1}M_{\mu}D_{21}^T)\bar{E}_2^{-1},\nonumber\\
C_c &=& \bar{E}_1^{-1}(B_2^T X + D_{12}^TM_{\tau}C_1)(I-Y X)^{-1}.
\end{eqnarray}

The following theorem, which is the main result of this paper, gives a sufficient condition for the
existence of a Lagrange stabilizing controller for the nonlinear system (\ref{Ch6sys3}),
(\ref{wi}), (\ref{nonlinearPeriod}), (\ref{Ch6sector_1}) :
\begin{theorem}\label{Ch6.OF.Theorem.B}  Suppose Assumptions   \ref{Assumption.A1}, \ref{Assump1},
 \ref{Ch6.Assum.Unobserv} and \ref{W.rational} hold for the nonlinear system (\ref{Ch6sys3}), (\ref{wi}), (\ref{nonlinearPeriod}), (\ref{Ch6sector_1}). Also, suppose
there exist constants $\lambda>0$ and $\tau_i>0,~i=0,\cdots,m$ such that the following conditions
are satisfied:
\begin{description}
\item[I.]\noindent The Riccati equation (\ref{Ch6ARE1}) has a stabilizing solution $X=X^T$ which is positive definite;
\item[II.]\noindent
The Riccati equation (\ref{Ch6ARE2}) has a pseudo-positive definite stabilizing solution $Y=Y^T$;
\item[III.]\noindent The matrix $XY$ has a spectral radius strictly less than one, $\rho(XY)<1$.
\end{description}
Then, 
the resulting closed-loop system corresponding to the controller (\ref{Ch6controller}),
(\ref{Ch6AREparameter.B}) is a pendulum-like system with respect to $\Pi(\tau_0\bar{p}\bar{d})$ and
is Lagrange stable.
 Here $\bar{p}=\rm{LCMD}(\nu)$.
\end{theorem}


%

\section{Output Feedback Lagrange Stabilizing Controller Synthesis for Uncontrollable  Systems } \label{sec:statefeedback} In this section, the state feedback and output feedback
pseudo $H_{\infty}$ control theories in Section \ref{sec:PseudoHinfinity} are applied to Lagrange
stabilization for nonlinear systems satisfying the following assumption which is dual to Assumption
\ref{Ch6.Assum.Unobserv}:
\begin{assumption}\label{Ch6vectorAssump}
There exists a non-zero vector $x$ such that $x^TA=0$ and $x^TB_2=0$.
\end{assumption}   

In a similar way to Assumption \ref{Ch6.Assum.Unobserv}, this assumption is also used to ensure
that the closed-loop system is pendulum-like and to rule out trivial cases in which the system can
be asymptotically stabilized. Also, this assumption implies that $(A,B_2)$ is not controllable.
Using the Kalman Decomposition \cite{Antsaklis2006}, it follows from Assumption
\ref{Ch6vectorAssump} that there exists a non-singular state-space transformation matrix $\bar{T}$
such that the matrices of the system (\ref{Ch6sys3}) are transformed to the form
 \begin{eqnarray} \label{Ch6.KD.a}\tilde{A}  &=& \bar{T}^{-1}AT=\left[
\begin{array}{cc}\tilde{A}_1 & \tilde{A}_{2}\\0 &
0\end{array}\right], ~\tilde{B}_2=\bar{T}^{-1}B_2=\left[\begin{array}{c}\tilde{B}_{2a}\\0\end{array}\right],\nonumber \\
\tilde{B}_1 &=&  \bar{T}^{-1} B_1 =  \left[\begin{array}{c}\tilde{B}_{1a} \\
 \tilde{B}_{1b}\end{array}\right],\nonumber\\ 
\tilde{D}_{12}  &=&  D_{12},
 \tilde{C}_1 = C_1\bar{T} =  \left[\begin{array}{cc}\tilde{C}_{1a}
 & \tilde{C}_{1b}\end{array}\right],\nonumber \\
\tilde{C}_{2} &=&  C_{2} \bar{T} = \left[\begin{array}{cc}\tilde{C}_{2a} &
\tilde{C}_{2b}\end{array}\right], ~\tilde{D}_{21}  =   D_{21},
\end{eqnarray} where $\tilde{A}_1\in\mathcal{R}^{(n-l)\times (n-l)}$, $\tilde{A}_2\in\mathcal{R}^{(n-l)\times l}$, $\tilde{B}_{2a}\in \mathcal{R}^{(n-l)\times q}$,
$\tilde{B}_{1a}\in \mathcal{R}^{(n-l)\times m}$, $\tilde{C}_{1a}, \tilde{C}_{2a}\in \mathcal{R}^{m\times (n-l)}$. 
\subsection{Output Feedback Lagrange Stabilization for Uncontrollable
 Systems}
The main result of this section involves the Riccati equations (\ref{Ch6ARE1}) and (\ref{Ch6ARE2}) which are dependent on parameters $\lambda>0$ and $\tau_i>0,~i=1,\cdots, m$. Using solutions $X$ and $Y$ to the equations (\ref{Ch6ARE1}) and (\ref{Ch6ARE2}) , we can construct the following matrices:
 \begin{eqnarray}\label{Ch5controllerP1} A_c &=& A+B_2C_c-B_cC_2\nonumber \\
&&+(B_1M_{\mu}M_{\tau}^{-1}M_{\mu}-B_cD_{21}M_{\mu}M_{\tau}^{-1}M_{\mu})B_1^TX,\nonumber\\
B_c &=& (I-YX)^{-1}(YC_2^T+B_1M_{\mu}M_{\tau}^{-1}M_{\mu}D_{21}^T)\bar{E}_2^{-1},\nonumber \\
C_c &=&
-\bar{E}_1^{-1}(B_2^TX+D_{12}^TM_{\tau}C_1). \end{eqnarray}

Also, we define two vectors of constants:
\begin{eqnarray} \label{Ch6.W.b}
\bar{d}_0&=&-\left[\begin{array}{cc}A_c & B_c \tilde{C}_{2a}\\\tilde{B}_{2a}C_c &
\tilde{A}_1\end{array}\right]^{-1} \left[\begin{array}{c}B_c\tilde{C}_{2b}\\
\tilde{A}_2\end{array}\right],\nonumber \\ 
\chi&=&\left[\begin{array}{cc}
-D_{12}\bar{E}_1^{-1}(B_2^TX+D_{12}^TM_{\tau}C_1) & C_1\end{array}\right]\bar{d},
\end{eqnarray}
where $ \bar{d}=\left[\begin{array}{cc}
                 I & 0 \\
                 0 & \bar{T}
               \end{array}
\right]\left[\begin{array}{c} \bar{d}_0\\ 1 \end{array} \right]$  with $\bar{T}$ defined in the
Kalman decomposition (\ref{Ch6.KD.a}). Using this notation, a sufficient condition for the solution
to the output feedback Lagrange stabilization Problem 1 can now be presented:
\begin{theorem}\label{Ch6.OF.Theorem.A}  Suppose Assumptions  \ref{Assumption.A1}, \ref{Assump1} and \ref{Ch6vectorAssump} hold for the system (\ref{Ch6sys3}), (\ref{wi}), (\ref{nonlinearPeriod}), (\ref{Ch6sector_1}). Also, suppose
there exist constants $\lambda>0$ and $\tau_i,~i=0,\cdots,m$ such that the following conditions are
satisfied for the nonlinear system  (\ref{Ch6sys3}), (\ref{wi}), (\ref{nonlinearPeriod}),
(\ref{Ch6sector_1}):
\begin{description}
\item[I.]\noindent The Riccati equation (\ref{Ch6ARE1}) has a stabilizing pseudo-positive definite solution $X=X^T$;
\item[II.]\noindent The Riccati equation (\ref{Ch6ARE2}) has a  stabilizing solution $Y=Y^T$ which
is positive definite;
\item[III.]\noindent The matrix $XY$ has a spectral radius strictly less than one, $\rho(XY)<1$;
\item[IV.] \noindent The matrix $ \left[\begin{array}{cc}A_c & B_c \tilde{C}_{2a}\\\tilde{B}_{2a}C_c &
\tilde{A}_1\end{array}\right]$ is non-singular and all the elements of the vector $
\nu=\tau_0\Delta^{-1}\chi$
are non-zero rational numbers, where $A_c$, $B_c$, $C_c$ and $\chi$ are defined in
(\ref{Ch5controllerP1}) and (\ref{Ch6.W.b}) using $X$, $Y$ in I, II and III.
\end{description}
Then, the closed-loop system consisting of  the system (\ref{Ch6sys3}), (\ref{wi}),
(\ref{nonlinearPeriod}), (\ref{Ch6sector_1}) and the controller (\ref{Ch6controller}),
(\ref{Ch5controllerP1}) is a pendulum-like system with respect to
$\Pi_{\lambda}=\{\bar{p}\tau_0\bar{d}\}$ and is Lagrange stable. Here $\bar{p}=\rm{LCMD}(\nu)$.
\end{theorem}

\subsection{Satisfaction of the rationality condition.} \label{Subsec:sufficient}
 Theorem \ref{Ch6.OF.Theorem.A} gives sufficient conditions for the existence of a solution to the Lagrange stabilizing controller synthesis problem for a nonlinear system satisfying
 Assumption \ref{Ch6vectorAssump}. However, the question arises as to whether, given $\lambda>0$, there will exist positive
constants $\tau_i,~0=1,\cdots,m,$ such that the stabilizing solutions to
 the Riccati equations (\ref{Ch6ARE1}) and (\ref{Ch6ARE2}) satisfy the rationality condition IV of this theorem.

First, we demonstrate that such $\tau=[\tau_1,\cdots,\tau_m]^T$, if exists,  can be constrained to
be a unit vector. Given any $\gamma> 0$, let $\hat{\tau}=\gamma \tau$,  $\tilde{X}=\gamma X$,
$\tilde{Y}=\gamma^{-1} Y$, $M_{\hat{\tau}}=\gamma M_{\tau}$, $\tilde{E}_1=D_{12}^T
M_{\hat{\tau}}D_{12}$
 and $\tilde{E}_2=D_{21}M_{\hat{\tau}}^{-1}D_{21}^T$. Multiplying  the
Riccati equation (\ref{Ch6ARE1}) by $\gamma$ and multiplying (\ref{Ch6ARE2}) by $\gamma^{-1}$ gives that
\begin{eqnarray}
\label{Ch6ARE1modified}
\lefteqn{(A+\lambda I -B_2\tilde{E}_1^{-1}D_{12}^T M_{\hat{\tau}} C_1)^T\tilde{X}+\tilde{X}(A+\lambda I-B_2\tilde{E}_1^{-1}D_{12}^T M_{\hat{\tau}}C_1)} \nonumber\\
&& +C_1^T(M_{\hat{\tau}}-M_{\hat{\tau}}D_{12}
\tilde{E}_1^{-1}D_{12}^TM_{\hat{\tau}})C_1\nonumber \\
&&+\tilde{X}({B}_1 M_{\mu} M_{\hat{\tau}}^{-1} M_{\mu}
{B}_1^T-B_2\tilde{E}_1^{-1}B_2^T)\tilde{X}=0, \\
\label{Ch6ARE2modified}
\lefteqn{(\lambda I+A-B_1M_{\mu}M_{\hat{\tau}}^{-1}M_{\mu} {D}_{21}^T\tilde{E}_2^{-1}C_2)\tilde{Y}}
\nonumber \\
 &&+ \tilde{Y}(\lambda I+A-B_1M_{\mu}M_{\hat{\tau}}^{-1}M_{\mu}{D}_{21}^T\tilde{E}_2^{-1}C_2)^T
\nonumber\\
&& +\tilde{Y}(C_1^TM_{\hat{\tau}}C_1-C_2^T\tilde{E}_2^{-1} C_2)\tilde{Y}\nonumber \\
&&+B_1
(M_{\mu}M_{\hat{\tau}}^{-1}M_{\mu}-M_{\mu}M_{\hat{\tau}}^{-1}D_{21}^T\tilde{E}_2^{-1}D_{21}M_{\hat{\tau}}^{-1}M_{\mu})B^T_1=0.\nonumber \\
\end{eqnarray}
It is obvious that (\ref{Ch6ARE1modified}) has the same form as (\ref{Ch6ARE1}) but both $X$ and
$M_{\tau}$ are scaled by $\gamma$. Also, (\ref{Ch6ARE2modified}) has the same form as
(\ref{Ch6ARE2}) but $Y$ is scaled by $\gamma^{-1}$ and $M_{\tau}$ is scaled by $\gamma$. Hence, Conditions I-III in the
statement of Theorem \ref{Ch6.OF.Theorem.A} are not affected if we use $\tilde{X}$, $\tilde{Y}$,
$M_{\hat{\tau}}$, $\tilde{E}_1$ and $\tilde{E}_2$ to replace $X$, $Y$, $M_{\tau}$, $\bar{E}_1$ and
$\bar{E}_2$ respectively. In addition, it is straightforward to verify that  Condition IV of
Theorem \ref{Ch6.OF.Theorem.A} is not affected by scaling the vector of constants
$\tau$. 
Thus, without loss of generality, we assume that  $\tau$ is a unit vector throughout the remainder
of this section, and if we take ${\tau}_i>0, ~i=1,\cdots,m-1$ as independent constants combined
into the vector $\bar{\tau}=[\tau_1, \cdots, \tau_{m-1}]$, then $\tau_m$ is given by
\begin{equation}\label{Ch6.Tau.m} \tau_m = \sqrt{1-\sum_{i=1}^{m-1} {\tau}_i^2}.\end{equation}


Define 
\[
\mathbb{T} = \left\{ \begin{array}{ll}{\bar{\tau}\in\mathcal{R}^{m-1}}: & \textmd{Equations (\ref{Ch6ARE1})
and (\ref{Ch6ARE2}) have}\\
&\textmd{nonsingular stabilizing solutions}\end{array}\right\}.
\]
 Let
$\tilde{\tau}=\left[\tau_0, \cdots, \tau_{m-1}\right]^T=[\tau_0,
  \bar{\tau}^T]^T$
  and define a function
$  f(\tilde{\tau})= \left[\begin{array}{ccc} f_1(\tilde{\tau}) &
 \cdots &  f_{m}(\tilde{\tau})
  \end{array}
\right]^T=\tau_0\Delta^{-1}\chi $ on the set
 $\mathbb{F}=\left\{\tilde{\tau}: \tau_0>0,
 \bar{\tau}\in\mathbb{T}\right\}.$ Let  $J(\tau_0, \tau_1,\cdots,\tau_{m-1})$ be the Jacobian matrix of $f(\tilde{\tau})$,
\begin{equation}\label{Ch6JacobianA}   J(\tau_0, \tau_1,\cdots,\tau_{m-1})=\left[\begin{array}{cccc}\frac{\partial f_1}{\partial \tau_{0}} & \frac{\partial
f_1}{\partial \tau_1} & \cdots & \frac{\partial f_1}{\partial \tau_{m-1}}  \\ \vdots & \vdots &
\ddots & \vdots \\ \frac{\partial f_m}{\partial \tau_{0}} & \frac{\partial f_m}{\partial \tau_1} &
\cdots & \frac{\partial f_m}{\partial \tau_{m-1}}
\end{array}\right].\end{equation} 
%
Then, we have $J(\tilde{\tau})=\Delta^{-1}\tilde{J}(\tilde{\tau})$ and the elements of
$\tilde{J}(\tilde{\tau})$ are
\begin{eqnarray}\label{Ch6.J.Tilde}
\tilde{J}_{i,1}  &=& w_i,\quad i=1,\cdots,m;\nonumber\\
\tilde{J}_{i,j} & =& \left\{
\begin{array}{l} \tau_0\left(\tau_i^{-2} w_i +  \tau_i^{-1} \frac{\partial w_i}{\partial \tau_i}\right):~ i=j,i=1,\cdots,m-1; \\
 \tau_0\tau_i^{-1} \frac{\partial w_i} {\partial \tau_j}:~i,j=1,\cdots,m-1, i\neq j;\end{array}\right.\nonumber \\
\tilde{J}_{m,j} & = & \tau_0\left(\tau_m^{-3}\tau_j w_m+\tau_m^{-1}\frac{\partial w_m} {\partial
\tau_j}\right), \quad j=1,\cdots,m.
\end{eqnarray}

The following theorem gives a sufficient condition for the existence of the constants
$\tau_0,\cdots,\tau_{m}$ satisfying all the conditions of Theorem \ref{Ch6.OF.Theorem.A}:
\begin{theorem}\label{Ch6RationalityTest}  Suppose Assumptions  \ref{Assumption.A1}, \ref{Assump1}  and \ref{Ch6vectorAssump} hold for the system (\ref{Ch6sys3}), (\ref{wi}), (\ref{nonlinearPeriod}), (\ref{Ch6sector_1}). Also, suppose
there exist a constant $\lambda>0$ and a vector of positive constants
$\tilde{\tau}=\left[\tau_0,\cdots,\tau_{m-1}\right]$ such that the following conditions are
satisfied for the system  (\ref{Ch6sys3}), (\ref{wi}), (\ref{nonlinearPeriod}),
(\ref{Ch6sector_1}):
\begin{description}
\item[(I)]\noindent Conditions I, II and III of Theorem
\ref{Ch6.OF.Theorem.A} hold;
\item[(II)] \noindent  $\det \tilde{J}(\tilde{\tau})\neq 0$ where the elements of
$\tilde{J}(\tilde{\tau})$ are defined as (\ref{Ch6.J.Tilde}).
\end{description}
Then, given any sufficiently small $\epsilon>0$, there exists
$\check{\tau}=\left[\check{\tau}_0,\check{\tau}_1,\cdots,\check{\tau}_{m-1}\right]\in\mathbb{F}$
such that $\|\check{\tau}-\tilde{\tau}\|<\epsilon$ and the constants $\tau_0=\check{\tau}_0$,
$\tau_i =\check{\tau}_i, i=1,\cdots, m-1$ and $\tau_m$ (defined as in (\ref{Ch6.Tau.m})) satisfy
all the conditions of Theorem \ref{Ch6.OF.Theorem.A} and hence the corresponding closed-loop system
is pendulum-like and Lagrange stable.
\end{theorem}

\section{State Feedback Lagrange Stabilization for Uncontrollable Systems}
In this section, we give a sufficient condition for the existence of a solution to the state
feedback Lagrange stabilization problem (Problem \ref{Problem2}) of Section II.

Using a solution to the Riccati equation (\ref{Ch6ARE1}), we define two vectors
$\bar{d}=\bar{T}[\bar{d}_0^T ~ 1]^T$ and $\chi=[ \chi_1 ~ \cdots ~ \chi_m]^T = \left(\left(I-
D_{12} \bar{E}_1^{-1}D_{12}^T M_{\tau}\right) C_1 - D_{12} \bar{E}_1^{-1}B_2^TX\right)\bar{d}$,
where $\bar{T}$ is defined by (\ref{Ch6.KD.a}) and
\begin{eqnarray*}
\bar{d}_0&=& - (\tilde{A}_1-\tilde{B}_{2a}\bar{E}_1^{-1} \tilde{D} _{12}^T M_{\tau} C_{1a} -
\tilde{B}_{2a} \bar{E}_1^{-1} \tilde{B}_{2a}^T
\bar{X}_{11})^{-1}\nonumber \\
&&\times (\tilde{A}_2-\tilde{B}_{2a}\bar{E}_1^{-1} \tilde{D} _{12}^T M_{\tau} C_{1b} -
\tilde{B}_{2a} \bar{E}_1^{-1} \tilde{B}_{2a}^T \bar{X}_{12})
\end{eqnarray*}
 with
~$\bar{X}_{11}\in\mathcal{R}^{(n-1)\times (m-1)}, ~~\bar{X}_{12}\in\mathcal{R}^{(n-1)\times 1}$
defined by $ \bar{T}^T X \bar{T}=\left[\begin{array}{cc}
            \bar{X}_{11} & \bar{X}_{12} \\
            \bar{X}_{12} & \bar{X}_{22}
          \end{array}
\right]$.
\begin{theorem}\label{Ch6StateFDsufficient}
Consider the nonlinear system (\ref{Ch6sys3a}), (\ref{Ch6sys3b}), (\ref{wi}),
(\ref{nonlinearPeriod}), (\ref{Ch6sector_1}) and suppose
Assumptions~ 
  \ref{Assumption.A1} and \ref{Ch6vectorAssump}
are satisfied.
 If there exist constants $\lambda>0$ and $\tau_i>0,~ i=0,\cdots,m$ such that the
Riccati equation (\ref{Ch6ARE1}) 
 has a pseudo-positive definite solution $X=X^T$  such that
\noindent\begin{description}\item[I.]\noindent  The matrix $A+\lambda I-B_2\bar{E}_1^{-1}D_{12}^T
M_{\tau} C_1+({B}_1 M_{\mu} M_{\tau}^{-1} M_{\mu} {B}_1^T -B_2\bar{E}_1^{-1}B_2^T)X$ is Hurwitz;
\noindent
\item[II.]  All elements of the vector $\nu=\tau_0\Delta^{-1}\chi$
are non-zero rational numbers.
  \end{description}
 Then, the closed-loop system corresponding to the state feedback control
\begin{equation}\label{Ch6StateFeedback}
u=\left(- D_{12} \bar{E}_1^{-1}D_{12}^T M_{\tau}C_1 - D_{12} \bar{E}_1^{-1}B_2^TP\right)
x\end{equation}
 is a pendulum-like system with respect to $\Pi\left(\bar{p}\tau_0\bar{d}\right)$ and is
Lagrange stable, where $\bar{p}=\rm{LCMD}(\nu)$.
\end{theorem}
%

 In a similar way to Theorem \ref{Ch6RationalityTest},  a sufficient condition for the existence of constants $\tau_0,\cdots,\tau_m$
satisfying Condition II of Theorem \ref{Ch6StateFDsufficient} is now given. The proof of this
result is similar to that of Theorem \ref{Ch6RationalityTest} and is omitted.
\begin{theorem}\label{Ch6RemoveRationalCondi} Consider the system (\ref{Ch6sys3a}), (\ref{Ch6sys3b}), (\ref{wi}),
(\ref{nonlinearPeriod}), (\ref{Ch6sector_1}) and suppose Assumptions~  \ref{Assumption.A1},
\ref{Ch6vectorAssump} are satisfied. Also, suppose there exists a constant $\lambda>0$ and a vector
of positive constants $\tilde{\tau}=\left[\tau_0,\cdots, \tau_{m-1}\right]^T$ satisfying the
following conditions:
  \begin{description} \item[I.] \noindent The Riccati equation
(\ref{Ch6ARE1}) has a pseudo-positive definite stabilizing solution $X$;
\item[II.] \noindent
$\tilde{J}\left(\tilde{\tau}\right)\neq 0$ where $\tilde{J}\left(\tilde{\tau}\right)$ is defined in
(\ref{Ch6.J.Tilde}).
\end{description}
Then, given any sufficiently small $\epsilon>0$, there exists a
$\check{\tau}=\left[\check{\tau}_0,\check{\tau}_1,\cdots,\check{\tau}_{m-1}\right]\in\mathbb{F}$
such that $\|\check{\tau}-\tilde{\tau}\|<\epsilon$ and the constants $\tau_0=\check{\tau}_0$,
$\tau_i=\check{\tau}_i, i=1,\cdots, m-1$ and $\tau_m$ (defined as in (\ref{Ch6.Tau.m})) satisfy all
the conditions of Theorem \ref{Ch6StateFDsufficient} and hence the corresponding closed-loop system
is pendulum-like and Lagrange stable.
\end{theorem}

\section{Illustrative Example}  \label{sec:Example}
\begin{figure}[htpb]
\begin{center}
 \psfrag{k1}{$k_1$} \psfrag{k2}{$k_2$} \psfrag{k3}{$k_3$} \psfrag{t1}{$\theta_1$}
\psfrag{t2}{$\theta_2$} \psfrag{t3}{$\theta_3$}
\includegraphics[width=7cm]{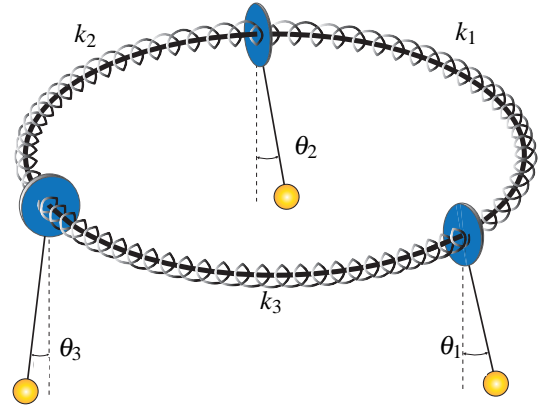}
 \caption{A system of three  pendulums connected on a ring by torsional springs.} \label{Ch6FigA0}
\end{center}
\end{figure}

To illustrate the theory developed in this paper, we consider a system consisting of three
connected pendulums, as shown in Figure \ref{Ch6FigA0}, where the pendulums are connected using
torsional springs and both pendulums and springs are supported by a rigid ring. The pendulums
oscillate in  planes perpendicular  to the ring and the torsional torque of the springs obeys the
angular form of the Hooke's law $F=-k \Delta \theta$, where $\Delta \theta$ is the angular
displacement, $F$ is the spring torque and $k$ is the torque constant. This system can be
considered as a prototype of many applications such as power systems, mechanical systems, network
systems, etc. Therefore, the Lagrange stabilization of  this system suggests many potential
applications of the proposed method. Suppose that the measurements consist of the angular velocity
of a pendulum and the angular difference between any two neighboring pendulums. As a result, all
absolute positions of the pendulums are unobservable. Also, our $A$ matrix has a single zero
eigenvalue which is an unobservable mode of the system. Hence, Assumption \ref{Ch6.Assum.Unobserv}
is satisfied. Let $x_1=\theta_1$, $x_2=\dot{\theta}_1$, $x_3=\theta_2$, $x_4=\dot{\theta}_2$,
$x_5=\theta_3$ and $x_6=\dot{\theta}_3$. Then, the system can be described by the state equations
of the form (\ref{Ch6sys3}) with the following matrices and nonlinearities
\begin{eqnarray*}
A &=& \left[\renewcommand{\arraystretch}{0.7} \begin{array}{cccccc}0 & 1 & 0 & 0 & 0 &
0\\ k_1+k_3 & -\alpha_1 & -k_1 & 0 & -k_3 & 0\\0 & 0 & 0 & 1 & 0 & 0\\-k_1 & 0 & k_1+k_2 &
-\alpha_2 & k_2 & 0\\0 & 0 & 0 & 0 & 0 & 1\\-k_3 & 0 & -k_2 & 0 & k_2+k_3 & -\alpha_3
\end{array}\right],\nonumber \\
 B_2&=&\left[ \renewcommand{\arraystretch}{0.7}\begin{array}{ccc}
                                           0 & 0 & 0\\
                                           1 & 0 & 0\\
                                           0 & 0 & 0\\
                                           0 & 1 & 0\\
                                           0 & 0 & 0 \\
                                           0 & 0 & 1
                                         \end{array}
\right],~ B_1=\beta\left[\renewcommand{\arraystretch}{0.7} \begin{array}{ccc}
                                           0 & 0 & 0\\
                                           1 & 0 & 0\\
                                           0 & 0 & 0\\
                                           0 & 1 & 0\\
                                           0 & 0 & 0 \\
                                           0 & 0 & 1
                                         \end{array}\right],\nonumber \\
 C_1 &=&  \left[\renewcommand{\arraystretch}{0.8} \begin{array}{cccccc}
                1 & 0 & 0 & 0 & 0 & 0\\
                0 & 0 & 1 & 0 & 0 & 0\\
                0 & 0 & 0 & 0 & 1 & 0
              \end{array}
\right],~  D_{12}= \epsilon_1 I_{3},\nonumber \\
 C_2&=&\gamma\left[\renewcommand{\arraystretch}{0.8}
\begin{array}{cccccc}
              1 & 0 & -1 & 0 & 0 & 0\\
              0 & 0 & 1 & 0 & -1  & 0\\
              0 & 0 & 0 & 0 & 0 & -1
            \end{array}
\right], \hspace{-.5cm}
\end{eqnarray*} 
\begin{equation} 
\label{syseqn} 
D_{21}=\epsilon_2 I_{3},\quad \textrm{and}\quad
\phi(z)=[\renewcommand{\arraystretch}{0.8}\begin{array}{ccc}
               \sin{z_1} & \sin{z_2} &  \sin{z_3}
             \end{array}
]^T.
\end{equation}

Note that this system has multiple nonlinearities and thus the results of
\cite{Wang2006,Yang2003,Li2005,Gao2009} cannot be applied. Also, the nonlinearities do not have
the special structure required in \cite{Ouyang08} to apply the result of that paper.

\begin{figure}[htpb]
\begin{minipage}{0.5\linewidth}
\begin{center}
\includegraphics[width=7.7cm]{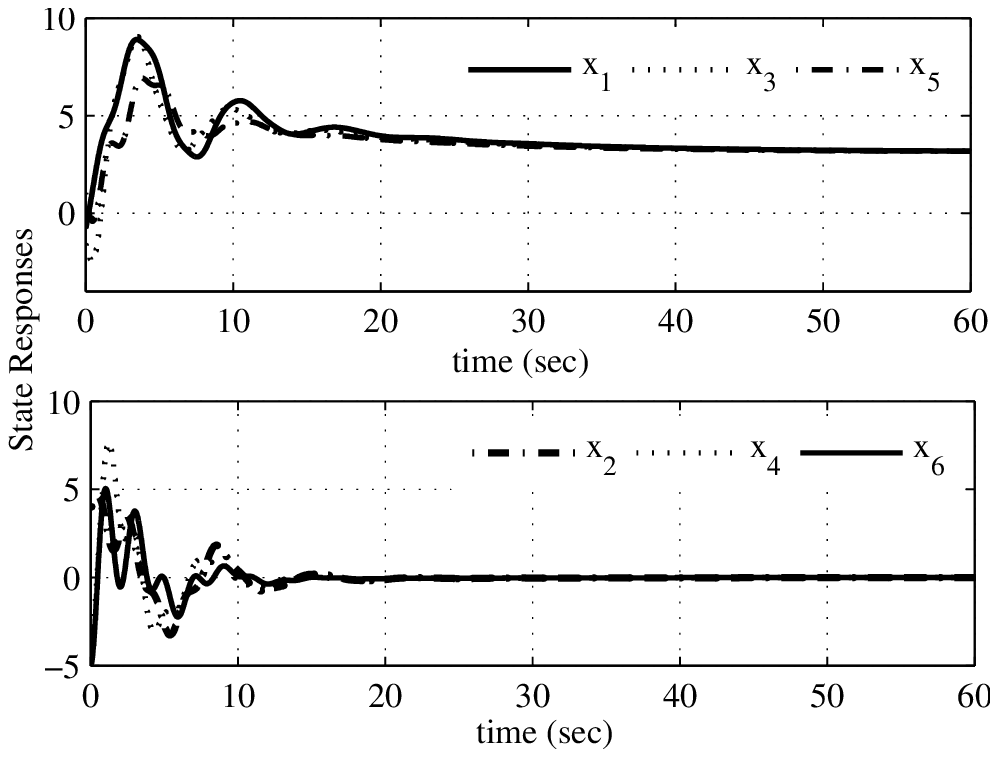}
\end{center}
\end{minipage}
\begin{minipage}{0.5\linewidth}
\begin{center}
\includegraphics[width=7.7cm]{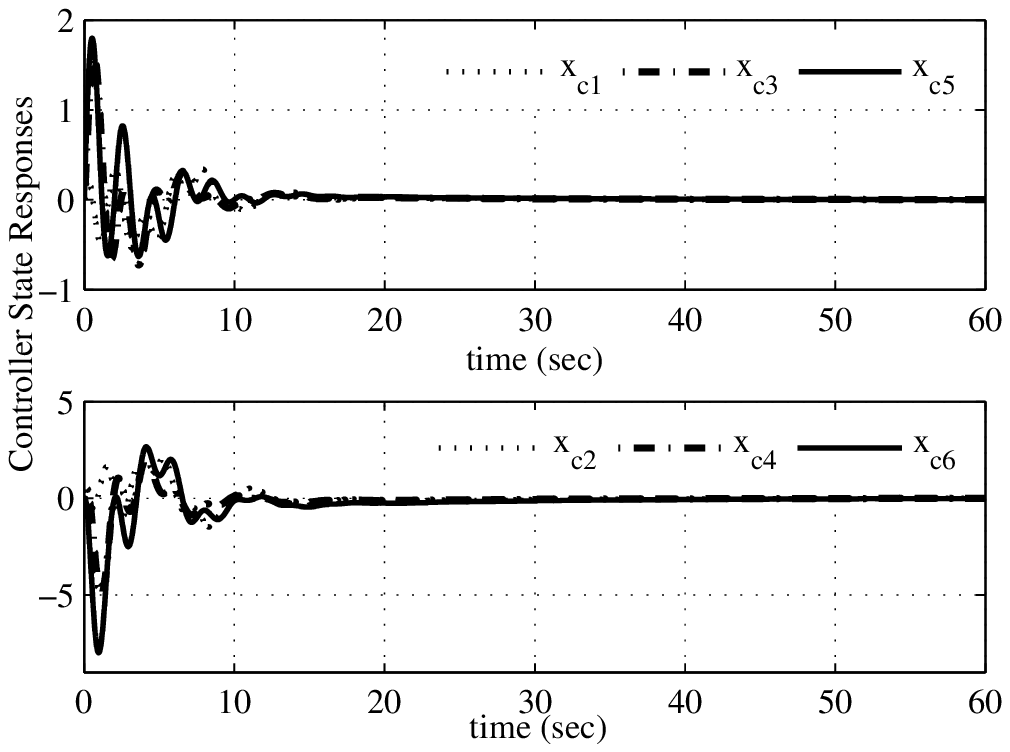}
\end{center}
\end{minipage}
\caption{System state responses and controller state responses of the closed-loop system for
initial values $[x_1(0),\cdots, x_6(0)]^T=[-\frac{\pi}{4}, 4, -\frac{\pi}{2}, -3, \frac{\pi}{3},
-5]$ and $x_{ci}=0, i=1,\cdots, 6$.} \label{Ch6FigB}
\end{figure}

 The damping coefficients are $\alpha_1=0.1$, $\alpha_2=0.05$,
$\alpha_3=0.08$. The torque constants are $k_1=0.02$, $k_2=0.03$, $k_3=0.05$. Also, we specify the
constants $\beta=0.2$, $\gamma=0.5$, $\epsilon_1=\epsilon_2=0.1$. It is easy to verify that the
system (\ref{Ch6sys3}), (\ref{syseqn}) satisfies Assumption \ref{Ch6.Assum.Unobserv}. Also, all of
the coefficients of the system (\ref{Ch6sys3}), (\ref{syseqn}) are rational. We choose
$\tau_0=2\pi$ to ensure that Assumption \ref{W.rational} is satisfied ($T$ will have rational
elements in this case). Therefore, Theorem \ref{Ch6.OF.Theorem.B} is applicable to the system.
Choosing $\tau_1=0.4$, $\tau_2=0.6$, $\tau_3=0.5$ and $\lambda=0.5$ and solving the Riccati
equations (\ref{Ch6ARE1}) and (\ref{Ch6ARE2}) gives solutions which satisfy all of the conditions
of Theorem \ref{Ch6.OF.Theorem.B}. Therefore, the solution to Problem 1 for the system
(\ref{Ch6sys3}), (\ref{syseqn}) can be constructed using this theorem. To illustrate the fact that
the resulting controller is such that the closed-loop system is Lagrange stable,
a series of simulations has been carried out with different initial values. These simulations have
confirmed that the trajectories of the closed-loop system are bounded. 
This can be seen in Figure \ref{Ch6FigB}, which shows the state responses of the system and the
controller state responses for one set of initial conditions, when the output feedback controller
is applied. { In addition, our simulations reveal that the trajectories of the closed-loop system
converge. Using Theorem 1 in \cite{Duan2007} and the results in \cite{leonov1996fdm}, it can be
verified that the closed-loop system has the property of dichotomy and the gradient-like property,
which explains the observed convergence.}

\section{Conclusions and Future Research}
This paper has studied the Lagrange stabilization problem for nonlinear systems with multiple
nonlinearities. In order to facilitate the controller synthesis for these systems, a
pseudo-$H_\infty$ control theory is developed. Sufficient conditions for the solution to state
feedback and output feedback pseudo-$H_\infty$ control problems are given. However, corresponding
necessary conditions are yet to be obtained.  The pseudo-$H_\infty$ control theory is applied to
solve output feedback and state  feedback Lagrange stabilization problems for nonlinear systems
with multiple nonlinearities. The efficacy of the method is illustrated by an example involving
coupled nonlinear pendulums on a ring.

This paper has considered the case where the nonlinear system contains decoupled nonlinearities. That is, as illustrated in Figure \ref{Ch6FigO}, we consider independent scalar nonlinearity blocks each subject to a sector bound constraint. One possible area for future research is to extend the approach of this paper to enable the consideration of nonlinear systems with coupled nonlinearities. This would involve allowing the nonlinear blocks in Figure \ref{Ch6FigO} to have vector inputs and outputs and to replace the sector bounds by more general local quadratic constraints.

\appendix
\subsection{Proof of Lemma \ref{rationalPendulum}}
First note that $p_i\neq 0$ since $\Delta_i\neq 0$. From the conditions of the lemma, we have
$E_i\bar{d}=\Delta_i \frac{q_i}{p_i}$. From (\ref{nonlinearPeriod}) and the fact that
$\frac{q_i}{p_i}\bar{p}$ is an integer, it follows that $\phi_i(t, C_{i} \bar{d} + C_{i} x)=
\phi_i(t, \Delta_i \frac{q_i}{p_i}\bar{p} + C_{i} x)=\phi_i(t,C_{i} x)$. As $A\bar{d}=0$, it
follows that,
\begin{equation}\label{DerivEq}A(x+\bar{d})+\sum_{i=1}^m B_{i} \phi_i(t, C_{i}\bar{d} + C_{i} x) =
Ax+\sum_{i=1}^m \phi_i(t, C_{i} x), \end{equation}  
for all $x$ and $t$.

Consider an arbitrary solution $x(t,t_0,x_0)$ of the system (\ref{Ch6sys0}), (\ref{wi}). Let
$\bar{x}(t)=x(t,t_0,x_0)+\bar{d}$ for $t\ge t_0$. Then, $\bar{x}(t_0)=x_0+\bar{d}$. Also, it
readily follows from (\ref{DerivEq}) that
$\bar{x}(t)=x(t,t_0,x_0+\bar{d})$. 
Furthermore, the local Lipschitz condition implies the uniqueness of this solution. Then, we have
$\bar{x}(t)=
x(t,t_0,x_0+\bar{d})=x(t,t_0,x_0)+\bar{d}$. 
 Hence, the lemma follows.              $\square$
\subsection{An Outline of the Proof  of Lemma \ref{LagrangeStability}:}
Define $\mathcal{G}\left( {x,\xi } \right)\dfn  \sum_{i=1}^m \tau _i \left( {-\mu
 _i^{ - 1} \xi_i -
C_{i} x } \right)^*   \left( {\mu _i^{ - 1} \xi_i - C_{i} x }
 \right)$ where $\xi\in \mathcal{C}^m$ and $x\in\mathcal{C}^n$ are arbitrary complex vectors.
Clearly, there exist constants $\delta>0$ and $0<\upsilon<1$ such that
\begin{eqnarray}\label{AugmentedFreq}
 \lefteqn{\left[ \begin{array}{c}\xi\\\zeta\end{array}\right]^* \left[\begin{array}{c}
    B^T \\ \left(\frac{\delta}{2\upsilon}\right)^{\frac{1}{2}} I
    \end{array}\right]} \nonumber \\
&& \times    \left((-j\omega-\lambda)I-A^T\right)^{-1}C^T M_{\tau}C\left((j\omega-\lambda)I-A\right)^{-1}\nonumber \\
&& \times \left[\begin{array}{c}
    B^T \\ \left(\frac{\delta}{2\upsilon}\right)^{\frac{1}{2}} I
    \end{array}\right]^T
\left[\begin{array}{c}\xi\\ \zeta\end{array}\right]
-\xi^* M_{\mu}^{-1}M_{\tau}M_{\mu}^{-1}\xi -\zeta^*
 \zeta \nonumber \\
&\le& -\frac{\delta}{2}\left(\xi^*
M_{\mu}^{-1}M_{\tau}M_{\mu}^{-1}\xi +\zeta^*
   \zeta\right),  \nonumber \\
&&\forall \omega\in \mathcal{R}, ~\forall [\xi^T~\zeta^T ]^T\in\mathcal{C}^{m+n}.
\end{eqnarray}

Given  $\omega\in\mathcal{R}$, we define 
$$ \bar{\sigma}=\left(\left(j\omega-\lambda\right)I-A\right)^{-1}[ B ~
\left(\frac{\delta}{2\upsilon}\right)^{\frac{1}{2}} I]\bar{\zeta}$$
 and $$ \mathcal{G}_a\left(\bar{\sigma},\bar{\zeta}\right)
  =\bar{\sigma}^* C^T M_{\tau} C
\bar{\sigma}-\bar{\zeta}^*M_a\bar{\zeta},$$
where $ \bar{\zeta}=\left[\begin{array}{c}\xi\\\zeta\end{array}\right]$ and $
M_a=\left[\begin{array}{cc} M_{\mu}^{-1}M_{\tau}M_{\mu}^{-1} & 0\\0 & I
\end{array}\right]$. Therefore, it
follows from (\ref{AugmentedFreq}) that
\begin{eqnarray}  \label{DrivedFreq}
  \mathcal{G}_a\left(\bar{\sigma},\bar{\zeta}\right)\le
    - \frac{\delta}{2}\bar{\zeta}^*M_a\bar{\zeta}, ~
\forall \omega\in\mathcal{R},  \bar{\zeta}\in\mathcal{C}^{m+n}.\end{eqnarray}
 Furthermore, since $M_a$ is a positive definite matrix, the inequality (\ref{DrivedFreq}) implies that
$\mathcal{G}_a\left(\bar{\sigma},\bar{\zeta}\right)<0,$
 for
all
 $\bar{\zeta}\in\mathcal{C}^{m+n}$ such
 that
$\|\bar{\zeta}\|\neq 0$. Also, the pair $(A,
 [B ~\sqrt{\frac{\delta}{2\upsilon}}I_{n\times
 n}])$ is controllable. Using
Theorem 1.11.1 in \cite{leonov1996fdm}, 
 it follows that there exists a Hermitian matrix
$P=P^*$ satisfying
$ 2x^*P((A+\lambda I)x+B\xi+\sqrt{\frac{\delta}{2\upsilon}}\zeta)+\bar{\sigma}^* C^T M_{\tau} C\bar{\sigma}
  -\xi^* M_{\mu}^{-1}M_{\tau}M_{\mu}^{-1}\xi-\zeta^* \zeta <0,$
for all $x\in \mathcal{C}^n$, $\bar{\zeta}=[\xi^T ~ \zeta^T]^T\in\mathcal{C}^{m+n}$ such
 that
$\|x\|+\|\xi\|+\|\zeta\|\neq 0$.
Letting $\zeta=0$, this implies that there exists an $n\times n$ matrix $P=P^T$ such that \begin{eqnarray}\label{DerivativeC} 2x^* P\left[A x +
B\xi\right]< -2 \lambda x^*Px -\mathcal{G}\left( {x,\xi} \right)
\end{eqnarray} for all $x\in \mathcal{C}^n$,
$\xi\in\mathcal{C}^m$ such that $\|x\|+\|\xi\|\neq 0$. Letting $\xi=0$ in (\ref{DerivativeC}),
we obtain 
that there exists a $r>0$ such that
$2x^T P\left[A+\lambda I\right]x <-r x^T x<0$.

Note that the pair $(A+\lambda I,r I)$ is observable. Since the matrix $A+\lambda I$ is
pseudo-Hurwitz, then using Theorem 3 in \cite{Chen1973} gives that $P$ is pseudo-positive definite.

In a similar way to the proof of Theorem 2.6.1 in \cite{leonov1996fdm}, we can prove that the set
$\{x\in\mathcal{R}^n: x^TPx<0\}$ is positively invariant for the nonlinear system (\ref{Ch6sys0}),
(\ref{wi}), (\ref{nonlinearPeriod}) and further prove that the solution $x(t,t_0,x_0)$ of the
system (\ref{Ch6sys0}), (\ref{wi}), (\ref{nonlinearPeriod}) is bounded. $\square$

\subsection{Proof of Theorem \ref{ESBRL0}:} In order to prove Theorem
\ref{ESBRL0}, some preliminary results are required. %
\begin{lemma}\label{Lagr_Lyap2} Suppose the pair $(C, A)$ has
no unobservable modes on the $j\omega$ axis. 
If the Lyapunov equation $A^TP+PA+C^TC=0$
has a pseudo-positive definite solution $P=P^T$ , then the matrix $A$ is
pseudo-Hurwitz.
\end{lemma}
In order to prove Lemma \ref{Lagr_Lyap2}, we require the following results:
\begin{lemma}[\cite{Jongen1987}]\label{Inertia_Schur} Let $\bar{P}$ be a symmetric
matrix of the form $\bar{P}=\left[ \begin{array}{cc}0 & \bar{P}_{12}\\ \bar{P}_{12}^T &
\bar{P}_{22}\end{array}\right]$, where $\bar{P}_{22}=\bar{P}_{22}^T$ and $\bar{P}_{12}^T$ are
$n_2\times n_2$ and $n_2\times n_1$ matrices, respectively. Also, let $k\dfn \mathrm{rank}
(\bar{P}_{12}^T)$. Then
\begin{equation}\mathrm{In}(\bar{P})=\mathrm{In}(\bar{P}_{22}/\mathrm{ker}(\bar{P}_{12}))+(k,k,
n_1-k), \end{equation} where $\mathrm{ker}(\bar{P}_{12})=\{\zeta\in
\mathbb{R}^{m}:\bar{P}_{12}\zeta=0\}$ and $\bar{P}_{22}/\mathrm{ker}(\bar{P}_{12})$ represents the
restriction of $\bar{P}_{22}$  to $\mathrm{ker}(\bar{P}_{12})$.
\end{lemma}

\begin{lemma}[\cite{Horn1985}]\label{eigenvectors_orth} If $A\in
  \mathcal{R}^{n\times n}$ and if
$\lambda, \mu\in\sigma(A)$ are eigenvalues of $A$ where $\lambda\neq\mu$, then
any left eigenvector of $A$ corresponding to $\mu$ is orthogonal to any right
eigenvector of $A$ corresponding to $\lambda$.
\end{lemma}

\emph{Proof of Lemma \ref{Lagr_Lyap2}:} The Kalman decomposition \cite{Antsaklis2006} establishes
the existence of a matrix $T$ which transforms the matrix pair $(A, C)$ into the form $
\bar{A}=TAT^{-1}=\left[\begin{array}{cc}\bar{A}_{11} & \bar{A}_{12}\\ 0 &
\bar{A}_{22}\end{array}\right]$, $\bar{C}=CT^{-1}= [0 ~ \bar{C}_2]$ where the pair $(\bar{A}_{22},
\bar{C}_2)$ is observable. The dimensions of the blocks in the above decomposition are as follows:
$\bar{A}_{11}\in \mathcal{R}^{n_1\times n_1}$, $\bar{A}_{11}\in \mathcal{R}^{n_1\times n_2}$,
$\bar{A}_{22}\in \mathcal{R}^{n_2\times n_2}$, and the column dimension of $\bar{C}_{2}$ is $n_2$.
Correspondingly,
let $ \bar{P}=T^{-T}PT^{-1}=\left[\begin{array}{cc}\bar{P}_{11} & \bar{P}_{12}\\
\bar{P}_{12}^T & \bar{P}_{22}\end{array}\right]$. It follows from the
observability of $(\bar{A}_{22}, \bar{C}_2)$ that there exists a matrix
$\bar{K}$ such that $\delta(\bar{A}+\bar{K}\bar{C} )=0$.

Using the equation $A^TP+PA+C^TC=0$, it follows that  %
%
\begin{eqnarray} \label{expand}\lefteqn{ \left[\begin{array}{c}\bar{A}_{11}^T\bar{P}_{11}
+\bar{ P } _ { 11 } \bar { A} _{11} \\
\bar{A}_{12}^T\bar{P}_{11}+\bar{A}_{22}^T\bar{P}_{12}^T +\bar{P}_{12}^T \bar{A}_{11}^T\end{array}\right.}\nonumber \\
&&\left.\begin{array}{c}\bar{P}_{11}\bar{A}_{12}+\bar{P}_{12}\bar{A}_{22}+\bar{A}_{11}\bar{P}_{12}
\\ \bar{P}_{12}^T\bar{A}_{12}+\bar{A}_{12}^T\bar{P}_
{12}+\bar{P}_{22}\bar{A}_{22}+\bar{A}_{22}^T\bar{P}_{22}+\bar{C}_2^T\bar{C}_2\end{array}
\right]=0.\nonumber \\
\end{eqnarray} Hence,
\begin{equation}\label{block11}\bar{A}_{11}^T\bar{P}_{11}+\bar{P}_{11}\bar{A}_{
11 } =0.
\end{equation}

\begin{claim}\label{claim}
If the pair $(\bar{C}, \bar{A})$ is such that there exists a matrix $\bar{K}$
satisfying $\delta(\bar{A}+\bar{K}\bar{C} )=0$, then $\mathrm{Re}
\lambda(\bar{A}_{11})\neq 0 ~\mathrm {for}~ \forall \lambda\in
\sigma(\bar{A}_{11})$.
\end{claim}

 To establish Claim \ref{claim}, we rewrite $\bar{K}$ as
  $ \bar{K}=\left[\begin{array}{c}\bar{K}_1\\\bar{K}_2\end{array}\right]$. Then $ \bar{A}+\bar{K}\bar{C}=\left[\begin{array}{cc} \bar{A}_{11} & \bar{A}_{12}+
\bar{K}_1\bar{C}_2\\
0 & \bar{A}_{22}+\bar{K}_2\bar{C}_2\end{array}\right]$. If there exists an
eigenvalue of $\bar{A}_{11}$ such that $\mathrm{Re}\lambda(\bar{A}_{11})=0$,
then $\bar{A}+\bar{K}\bar{C}$ obviously has purely imaginary eigenvalues. This
contradicts the fact that $\bar{K}$ is chosen so that
$\delta(\bar{A}+\bar{K}\bar{C} )=0$. Therefore,
$\mathrm{Re}\lambda(\bar{A}_{11})\neq 0$. This completes the proof of the
claim.

Combining Claim \ref{claim} and (\ref{block11}) gives that
$\bar{P}_{11}=0$. 
Also, the $(1, 2)$ block of
(\ref{expand}) implies that
\begin{equation}\label{block12}
\bar{P}_{12}\bar{A}_{22}+\bar{A}_{11}\bar{P}_{12} =0.
 \end{equation}

As $\bar{P}$ is nonsingular, this implies that $\bar{P}_{12}\bar{P}_{12}^T>0$.
Applying Lemma \ref{Inertia_Schur}  to $\bar{P}$ gives that
$\mathrm{In}~\bar{P}=\mathrm{In}\left(\bar{P}_{22}/\mathrm{ker}
\bar{P}_{12}\right)+(\mathrm{rank}\bar{P}_{12}^T,\mathrm{rank}\bar{P}_{12}^T,
n_1-\mathrm{rank}\bar{P}_{12}^T)$. It is known that $\mathrm{In}\bar{P}=\left(n-1,1,0\right)$. This
implies that
\begin{enumerate}
  \item $0=\delta(\bar{P})=\delta\left(\bar{P}_{22}/\mathrm{ker}
\bar{P}_{12}\right)+(n_1-\mathrm{rank}\bar{P}_{12}^T)$. For
$\bar{P}_{12}^T\in\mathcal{R}^{n_2\times n_1}$, it always holds that
$\mathrm{rank}\bar{P}_{12}^T\le n_1$. Then,
$\delta\left(\bar{P}_{22}/\mathrm{ker} \bar{P}_{12}\right)\le 0$. So,
$\delta\left(\bar{P}_{22}/\mathrm{ker} \bar{P}_{12}\right)= 0$ holds. This
further implies that $n_1=\mathrm{rank}\bar{P}_{12}^T$. Also, the condition
$\delta\left(\bar{P}_{22}/\mathrm{ker} \bar{P}_{12}\right)=0$ implies that the
matrix $\bar{P}_{22}/\mathrm{ker} \bar{P}_{12}$ is nonsingular and has no
purely imaginary eigenvalues.
  \item $1=\nu(\bar{P})=\nu\left(\bar{P}_{22}/\mathrm{ker}
\bar{P}_{12}\right)+\mathrm{rank}\bar{P}_{12}^T$. Hence, $\mathrm{rank}\bar{P}_{12}^T\le 1$ and
$\bar{P}_{12}\bar{P}_{12}^T>0$ imply $\mathrm{rank}\bar{P}_{12}^T=1$ and
$\nu\left(\bar{P}_{22}/\mathrm{ker} \bar{P}_{12}\right)=0$. Hence, $\bar{P}_{22}/\mathrm{ker}
\bar{P}_{12}\dfn S^T\bar{P}_{22}S$ is symmetric and positive definite, where the columns of $S$
form a basis for $\mathrm{ker}\bar{P}_{12}$.
  \item Finally, the identity $\pi(\bar{P})=n-1=\pi\left(\bar{P}_{22}/\mathrm{ker}
  \bar{P}_{12}\right)+\mathrm{rank}\bar{P}_{12}^T$ implies $\pi\left
(\bar{P}_{22}/\mathrm{ker}
  \bar{P}_{12}\right)=n-2.$
\end{enumerate}
Therefore, it follows that $n_1=1$. Hence $A_{11}$ is a scalar, $\bar{P}_{12}$
is a row vector of dimension $n-1$, and $\bar{P}_{22}$ is a $(n-1)\times (n-1)$
matrix. The dimension of $S^T\bar{P}_{22}S$ is equal to $n-2$.

As $\bar{P}_{12}\in \mathcal{R}^{1\times (n-1)}$, $\bar{P}_{12}\neq 0$,
(\ref{block12}) implies that $-\bar{A}_{11}$ is an eigenvalue of $\bar{A}_{2
2}$ and $\bar{P}_{12}$ is the corresponding left eigenvector.
 Now, let $\lambda$ be any eigenvalue of $\bar{A}_{22}$ such that $\lambda\neq
-\bar{A}_{11}$ and let $q$ be a corresponding right eigenvector;  that is, $\bar{A}_{22}q=\lambda
q$. Then Lemma \ref{eigenvectors_orth} implies that $\bar{P}_{12}q=0$. Hence, $q\in
\mathrm{Ker}~\bar{P}_{12}$.

Pre- and post-multiplying the $(2,2)$ block of (\ref{expand}) by $q^T$ and $q$ respectively implies
that $q^T\bar{P}_{22}\bar{A}_{22}q+q^T\bar{A}_{22}^T\bar{P}_{22}q+q^T\bar{C}_2^T \bar{C}_2q=0$.
Therefore, $\lambda q^T\bar{P}_{22}q+\bar{\lambda}q^T\bar{P}_{22}q+\| \bar{C}_2q\|^2=0$.

Using the fact that $\alpha=q^T\bar{P}_{22}q$ is positive on $\mathrm{Ker} \bar{P}_{12}$, we have
$2\alpha \mathrm{R e}(\lambda)+\|\bar{C}_2q\|^2=0$. Since $q\neq 0$ is an eigenvector of
$\bar{A}_{22}$, $\bar{C}_2 q\neq 0$. Therefore,
 $\mathrm{Re}(\lambda)<0$.

The above derivation shows that all eigenvalues of $\bar{A}_{22}$, possibly with the exception of
$-\bar{A}_{11}$, have negative real part. Therefore, if $-\bar{A}_{11}$ is negative, then
 $\bar{A}_{22}$ is Hurwitz; if $-\bar{A}_{11}$ is positive, then
 $\bar{A}_{22}$ has all the eigenvalues $\lambda\neq -\bar{A}_{11}$ negative
except  $-\bar{A}_{11}$.

Now, we can conclude that the spectrum of $\bar{A}$ is
$\sigma(\bar{A})=\{-\bar{A}_{11},\bar{A}_{11} ~\mathrm{and}~ \lambda:\lambda\neq -\bar{A}_{11},
\mathrm{Re} \lambda<0\}$. Also, since the pair $(C,A)$ has no unobservable modes on the imaginary
axis, it follows that $\bar{A}_{11}\neq 0$. Hence, $\bar{A}$ is pseudo-Hurwitz. This completes the
proof of Lemma \ref{Lagr_Lyap2}. $\square$

\emph{Proof of Theorem \ref{ESBRL0}}: By assumption, $P$ is such that $\delta(A+BB^TP)=0$. Letting
$\bar{C}=\left[ \begin{array}{c}B^TP\\C\end{array}\right]$, $ \bar{K}=\left[\begin{array}{cc}B &
0\end{array}\right]$,  it follows that $A+\bar{K}\bar{C}$ is such that
$\delta(A+\bar{K}\bar{C})=0$. Therefore, $(A+\bar{K}\bar{C}, \bar{C})$ has no unobservable mode on
the imaginary axis and hence $(A, \bar{C})$ has no unobservable mode on the imaginary axis, either.
Applying Lemma \ref{Lagr_Lyap2} to the Lyapunov equation $A^TP+PA+\bar {C}^T\bar{C}=0$, it follows
that $A$ is pseudo-Hurwitz. Hence, $\det(j\omega I-A)\neq 0$ for all $\omega\in\mathcal{R}$.

Now, we show that (\ref{VO.ppr}) holds. Since $A$ is pseudo-Hurwitz, then $\det (j\omega -A)\neq
0$, $\forall \omega\in\mathcal{R}$. Hence, (\ref{ARE0}) implies that
\begin{eqnarray}\label{Frequency_equality} 
\lefteqn{G(-j\omega)^T
G(j\omega)=} \nonumber \\
&&I- [I-B^TP(-j\omega I-A)^{-1}B]^T[I-B^TP(j\omega I-A)^{-1}B] \nonumber \\
&&\leq I
\end{eqnarray} for all $\omega\ge 0$. It follows that $\max\limits_{\omega\in \mathcal{R}}
\{\sigma_{max}[G(j\omega)G(-j\omega)^T]\}\le 1$. Furthermore, note that $G(j\omega)\rightarrow 0$
as $\omega\rightarrow \infty$. Now suppose that there exists an $\bar{\omega}\ge 0$ such that
$\max\limits_\omega\{ \sigma_{max}[G(-j\omega)^T G(j\omega)]\}=1$. It follows from
(\ref{Frequency_equality}) that there exists a vector $z$ such that
$[I-B^TP(j\bar{\omega}-A)^{-1}B]z=0$. Hence, $\det[I-B^TP(j\bar{\omega}-A)^{-1}B]=0$. However,
using a standard result on determinants, it follows that $\det[j\bar{\omega}I-A-BB^TP]
=\det[j\bar{\omega}I-A]\det[I-B^TP(j\bar{\omega}I-A)^{-1}B]$.
Thus $\det[j\bar{\omega}I-A-BB^TP]=0$. This  conclusion contradicts the assumption that
$\delta(A+BB^TP)=0$. Hence, (\ref{VO.ppr}) holds.
$\square$
\subsection{Proof of Theorem \ref{ESBRL00}:}
%
%
 Let $\mu\dfn \left(\max\limits_{\omega\in\mathcal{R}}\{ \sigma_{max} [(-j\omega I-A^T)^{-1}C^TC(j\omega I-A)^{-1}]\}\right)^{\frac{1}{2}}$.
 It follows from (\ref{VO.ppr}) that there exist an $\epsilon\ge 0$  such that
 $G(j\omega)G(j\omega)^T\le (1-\epsilon)I$.
 Hence, $\frac{\epsilon}{2\mu^2}C(j\omega I -A)^{-1}
 (-j\omega I-A^T)^{-1}C^T\le\frac{\epsilon}{2}I$
 for all $\omega\ge 0$. Then, given
 any $\omega\ge 0$, $C(j\omega I -A)^{-1}\tilde{B}\tilde{B}^T(-j\omega I-A^T)^{-1}C^T\le
 \left(1-\frac{\epsilon}{2}\right)I$,
 where $\tilde{B}$ is a non-singular matrix defined by $\tilde{B}\tilde{B}^T=BB^T+\epsilon/2\mu^2 I$.
 This further implies that
 \begin{equation}\label{Trf_Ineq40} \tilde{B}^T(-j\omega I-A^T)^{-1}C^TC(j\omega I -A)^{-1}\tilde{B}\le
 \left(1-\frac{\epsilon}{2}\right)I\end{equation}
 for all $\omega\ge 0$. Let $\eta^2 \dfn \max\limits_{\omega\in\mathcal{R}}\sigma_{max} [\tilde{B}^T(-j\omega I-A^T)^{-1}C^TC(j\omega I -A)^{-1}\tilde{B}]$.
 Hence, $\frac{\epsilon}{2\eta^2}\tilde{B}^T(-j\omega
 I-A^T)^{-1}(j\omega I -A)^{-1}\tilde{B}\le \frac{\epsilon}{2} I$,
 holds for all $\omega\ge 0$. From (\ref{Trf_Ineq40}), it follows that
 given any $\omega\ge 0$,
 \begin{equation}\label{frequency_tilde}\tilde{G}(-j\omega)^T\tilde{G}(j\omega)\le
 I\end{equation}
 where $\tilde{G}(s)=\tilde{C}(sI-A)^{-1}\tilde{B}$ with  $\tilde{C}$ being a non-singular matrix defined so that $\tilde{C}^T\tilde{C}=C^TC+(\epsilon/2\eta^2)
 I$. Furthermore, (\ref{frequency_tilde}) implies $\tilde{G}(j\omega)\tilde{G}(-j\omega)^T \le
 I$.


 Since $A$ has no eigenvalue on the $j\omega$-axis and the pair $(A, \tilde{B})$ is stabilizable
  (since it is controllable), it follows from
Theorem 13.34 in \cite{Zhou1996} and (\ref{frequency_tilde}) that there exists a right coprime
factorization $ \tilde{G}(s) = \tilde{N}(s)\tilde{M}^{ - 1}(s)$  such that $\tilde{M}(s)\in
\mathcal{RH}_\infty$ is an inner transfer function matrix where $\tilde{M}(s)
=\tilde{F}(sI-A-\tilde{B}\tilde{F})^{-1}\tilde{B}+I \in \mathcal{RH}_\infty$,
$\tilde{N}(s)=\tilde{C}(sI-A-\tilde{B}\tilde{F})^{-1}\tilde{B}\in \mathcal{RH}_\infty$ with
$\tilde{F}=-\tilde{B}^{T}\tilde{X}$, and the Riccati equation $A^{T}\tilde{X} + \tilde{X}A -
\tilde{X}\tilde{B}\tilde{B}^{T}\tilde{X} = 0$
has a solution $\tilde{X} \ge 0$ such that $A-\tilde{B}\tilde{B}^T\tilde{X}$ is stable. Since
$\tilde{M}(s)$ is an inner transfer function, it follows that
$\tilde{N}(j\omega)\tilde{N}^T(-j\omega)=\tilde{G}(j\omega)\tilde{G}^T(-j\omega)\le I$. Applying
the bounded real lemma (e.g., see \cite{Anderson1973}), the above condition is equivalent to the
existence of a stabilizing solution to the Riccati equation
\begin{eqnarray}  \label{AREx}
(A-\tilde{B}\tilde{B}^T\tilde{X})^T\hat{P}+\hat{P}(A-\tilde{B}\tilde{B}^T\tilde{X})
+\hat{P}\tilde{B}\tilde{B}^{T}\hat{P}+\tilde{C}^{T}\tilde{C}=0.
\end{eqnarray}

Let $\tilde{P}=\hat{P}-\tilde{X}$. Then substituting this into (\ref{AREx})
gives that
\begin{eqnarray}\label{ARE_Px} \lefteqn{( A^T \tilde{X} + \tilde{X} A -\tilde{X}^T \tilde{B}\tilde{B}^T \tilde{X})} \nonumber \\
&&+ (A^T \tilde{P} +
\tilde{P}A + \tilde{P}\tilde{B}\tilde{B}^T \tilde{P} + \tilde{C}^{T}\tilde{C})=0.
\end{eqnarray}
Therefore, (\ref{ARE_Px}) implies that $ A^T \tilde{P} + \tilde{P}A + \tilde{P}BB^T \tilde{P} +
C^{T}C+\frac{\epsilon}{2\mu^2}\tilde{P}^2+ \frac{\epsilon}{2\eta^2} I =0$. This implies that $P=\tilde{P}$ satisfies (\ref{Riccati_inequality}). 
This proves the first claim of the theorem. Now we prove the second claim.

From (\ref{VO.ppr}), it follows that $G(j\omega)G^{T}(-j\omega)\le I$. As the pair $(A,B)$ is
stabilizable, Theorem 13.34 in \cite{Zhou1996} implies that there exists a right coprime
factorization $ G(s) = N(s)M^{ - 1}(s)$ such that $M(s)\in \mathcal{RH}_\infty$ is an inner
transfer function matrix where $M(s) =F(sI-A-BF)^{-1}B+I \in \mathcal{RH}_\infty$,
$N(s)=C(sI-A-BF)^{-1}B\in \mathcal{RH}_\infty$ with $F=-B^{T}X$, and the  Riccati equation $A^{T}X
+ XA - XBB^{T}X = 0$
has a solution $X \ge 0$ such that $A-BB^TX$ is stable. Since $M(s)$ is an inner transfer function,
it follows that $N(j\omega)N^T(-j\omega)=G(j\omega)G^T(-j\omega)\le I$. Applying the bounded real
lemma \cite{Anderson1973}, the above condition is equivalent to the condition that the following
Riccati equation has a stabilizing solution
\begin{eqnarray}
(A-BB^TX)^T\bar{P}+\bar{P}(A-BB^TX) +\bar{P}BB^{T}\bar{P}+C^{T}C=0.
\end{eqnarray}

Let $P=\bar{P}-X$. Then substituting this into (\ref{AREx0}) gives that
\begin{eqnarray}\label{AREx0}(A^T X + X A  -   X^T BB^T X) +  (A^T P + PA + PBB^T P + C^{T}C) = 0.
\end{eqnarray}
Therefore, the Riccati equation (\ref{ARE0}) has a stabilizing solution. Furthermore, as the pair
$(A, C)$ is observable, it follows from the Inertia theorem in \cite{Chen1975} that the solution
$P=P^T$ of the Riccati equation (\ref{ARE0}) is a pseudo-positive definite matrix. This completes
the proof. $\square$

\emph{Proof of Theorem \ref{synthesis0}: }
The Riccati equation (\ref{H_controller}) can be written as
 \begin{eqnarray} \label{ARE_plus0}
\lefteqn{(A-B_2E_1^{-1}D_{12}^TC_1)^TP+P(A-B_2E_1^{-1}D_{12}^TC_1)}\nonumber\\
&&+PB_1B_1^TP+ C_1^T(I-D_{12}E_1^{-1}D_{12}^T)C_1\nonumber \\
&&-PB_2E_1^{-1}D_{12}^T(I-D_{12}E_1^{-1}D_{12}^T)C_1- PB_2E_1^{-1}B_2^TP\nonumber \\
&&-C_1^T(I-D_{12}E_1^{-1}D_{12}^T)D_{12}E_1^{-1}B_2^TP=0.
 \end{eqnarray}
As the Riccati equation (\ref{H_controller}) has a solution $P=P^T$ which is pseudo-positive
definite, the equation (\ref{ARE_plus0}) also has this property. Substituting
$K=-E_1^{-1}\left(D_{12}^TC_1+B_2^TP\right)$  into (\ref{ARE_plus0}) implies that
\begin{eqnarray}\label{ARE_control}
\lefteqn{(A+B_2K)^TP+P(A+B_2K)+PB_1B_1^TP}\nonumber \\
&&+(C_1+D_{12}K)^T(C_1+D_{12}K)=0
\end{eqnarray}
has a solution $P=P^T$ which is pseudo-positive definite. Also, the fact that
the matrix (\ref{stabilizingmatrix}) has no purely imaginary eigenvalues implies that
$A+B_2K+B_1B_1^TP$ has no purely imaginary eigenvalues. Therefore, it follows from Theorem
\ref{ESBRL0} that the resulting closed-loop system
\begin{eqnarray*}\dot{x} &=& \left(A-B_2E_1^{-1}D_{12}^TC_1-B_2E_1^{-1}B_2^TP\right)x+B_1w,\\
                 z &=& C_1-D_{12}E_1^{-1}\left(D_{12}^TC_1+B_2^TP\right)x\end{eqnarray*} is pseudo strict bounded real. 
 This completes the proof of Theorem \ref{synthesis0}.  $\square$
\subsection{Proof of Theorem \ref{theorem20}}
In order to prove Theorem \ref{theorem20}, the following lemma is introduced.
\begin{lemma}\label{lemma1}Suppose the conditions  of Theorem \ref{theorem20} hold.
Then, the matrix $Z\dfn \left(I-YX\right)^{-1}Y=Y\left(I-XY\right)^{-1}> 0$ is
a stabilizing solution to the Riccati equation \begin{equation}\label{AREz}
A_*Z+ZA_*^T-ZM_*Z+N_*=0
\end{equation} where $A_*= A-B_1D_{21}^TE_2^{-1}C_2+ B_1(I-D_{21}^TE_2^{-1}D_{21})B_1^TX$, $N_*=B_1(I-D_{21}^TE_2^{-1}D_{21}B_1^{-1})$,
$M_*=(C_2+D_{21}B_1^TX)^TE_2^{-1}(C_2+D_{21}B_1^TX)
   -(B_2^TX+D_{12}^TC_1)^TE_1^{-1}(B_2^TX+D_{12}^TC_1)$.
\end{lemma}
%
%
The proof of this lemma is similar to that of Lemma 3.2 in \cite{Petersen1991} and  is omitted.

 \emph{Proof of Theorem \ref{theorem20}}: We will prove that the
compensator of the form (\ref{Ch6controller}), (\ref{conParameters})
makes the closed-loop system pseudo strict bounded real. In order to establish this fact, note that
 Lemma \ref{lemma1} implies that matrix $Z=(I-YX)^{-1}Y> 0$ is a stabilizing solution to the Riccati equation (\ref{AREz}).
 Substituting $(I-YX)^{-1}Y=Z$ and $(I-YX)^{-1}=(I+ZX)$ into (\ref{conParameters}), it follows that the compensator
 input matrix $B_c$ can be written as \begin{equation}\label{Bc} B_c=B_1D_{21}^TE_2^{-1}+Z(C_2^T+XB_1D_{21}^T)E_2^{-1}.\end{equation}

 We now form the closed-loop system associated with system (\ref{Ch6sys3}) and compensator
 (\ref{Ch6controller}). This system is described by the state equation
 \begin{eqnarray}\label{AugmentedSys} \dot{\eta}&=&\bar{A}\eta+\bar{B}w,\nonumber\\
 z&=&\bar{C}\eta,\end{eqnarray}
 where  $\eta\dfn \left[ \begin{array}{c}x\\x-x_c\end{array}\right]$,~ $\bar{A}\dfn \left[ \begin{array}{cc}
 A+B_2C_c & -B_2C_c\\A-A_c+B_2C_c-B_cC_2 & A_c-B_2C_c\end{array}\right]$,~ $\bar{B}\dfn \left[ \begin{array}{c}B_1\\B_1-B_cD_{21}\end{array}\right]$
 and $ \bar{C}\dfn \left[\begin{array}{cc} C_1+D_{12}C_c & -D_{12}C_c\end{array}\right]$.

 In order to verify that this system is pseudo strict bounded real, we first recall that $Z> 0$ is a
 stabilizing solution to the Riccati equation (\ref{AREz}). This implies that $Z>0$ will also be a
 stabilizing solution to the Riccati equation
 \begin{equation}\label{ARE00}A_0Z+ZA_0^T+ZC_0^TC_0Z+B_0B_0^T=0\end{equation}
 where $ A_0  \dfn  A-B_1D_{21}^TE_2^{-1}C_2+ B_1(I-D_{21}^TE_2^{-1}D_{21})B_1^TX- Z(C_2+D_{21}B_1^TX)^TE_2^{-1}(C_2+D_{21}B_1^TX)$,
 $B_0 \dfn B_1 (I-D_{21}^T E_2^{-1} D_{21}) - Z(C_2+D_{21}B_1^T X)^T E_2^{-1}D_{21}$, $C_0  \dfn  E_1^{\frac{1}{2}} (B_2^T X + D_{12}^T C_1)$.

Let $W=Z^{-1}>0$, then the Riccati equation (\ref{ARE00}) leads to
\begin{equation}\label{AREw}A_0^TW+WA_0+WB_0B_0^TW+C_0^TC_0=0.\end{equation}
Now, we prove that $W=W^T$  is an anti-stabilizing solution of (\ref{AREw}). Using the Riccati
equation (\ref{AREw}), it follows that $-(A_0+ZC_0^TC_0)=Z(A_0^T+Z^{-1}B_0B_0^T)Z^{-1}$.  Hence the
matrix $-(A_0^T+ZC_0^TC_0)$ is similar to the matrix $(A_0+B_0B_0^TW)^T$. Since $Z$ is a
stabilizing solution to (\ref{ARE00}), the matrix $A_0^T+ZC_0^TC_0$ must be Hurwitz and hence the
matrix $A_0+ B_0B_0^TW$ must be anti-Hurwitz; i.e., $W$ is an anti-stabilizing solution to
(\ref{AREw}).

 Now, we define $\Sigma\dfn \left[ \begin{array}{cc} X & 0\\
0 & W\end{array}\right].$ As $X$ is pseudo-positive definite and $W>0$, it follows that $\Sigma$ is
also pseudo-positive definite. Using equations (\ref{AREp1}), (\ref{conParameters}), (\ref{Bc}),
(\ref{AREw}), it is straightforward to verify that $\Sigma$ satisfies the Riccati equation $
\bar{A}^T\Sigma+\Sigma \bar{A}+\Sigma\bar{B}\bar{B}^T\Sigma+\bar{C}^T\bar{C}=0$.
Furthermore, it is straightforward to verify that $
\bar{A}+\bar{B}\bar{B}^T\Sigma=\left[\begin{array}{cc}\check{A}_{11}& \check{A}_{12}\\0 &
A_0+B_0B_0^TW\end{array}\right]$
where $ \check{A}_{11}=A-B_2E_1^{-1}D_{12}^TC_1-(B_2E_1^{-1}B_2^T-B_1B_1^T)X$, $
\check{A}_{12}=B_2E_1^{-1}B_2^T+B_2E_1^{-1}D_{12}^TC_1+B_1(I-D_{21}^TE_2^{-1}D_{21})B_1^TW-B_1D_{21}^TE_2^{-1}(C_2+D_{21}^T)ZW$.
Using the fact that $X$ is a stabilizing solution to (\ref{AREp1}) and $W$ is an anti-stabilizing
solution to (\ref{AREw}), it follows that $\bar{A}+\bar{B}\bar{B}^T\Sigma$ has no purely imaginary
eigenvalues. We have noted previously that the matrix $\Sigma$ is pseudo-positive definite.
Therefore, using Theorem \ref{ESBRL0}, we conclude that the system (\ref{AugmentedSys}) is pseudo strict bounded real. Using the fact that  $\eta=\left[ \begin{array}{cc}I & 0\\
I & -I\end{array}\right]\left[ \begin{array}{c}x\\x_c\end{array}\right]$, it follows that the
closed-loop system
 \begin{eqnarray*}\left[ \begin{array}{c}\dot{x}\\\dot{x}_c\end{array}\right]&=&\left[ \begin{array}{cc}
 A & B_2C_c\\B_cC_2 & A_c\end{array}\right]\left[ \begin{array}{c}x\\x_c\end{array}\right]+\left[ \begin{array}{c}B_1\\B_cD_{21}\end{array}\right]w;\nonumber\\
 z&=&\left[ \begin{array}{cc} C_1 & D_{12}C_c\end{array}\right]\left[ \begin{array}{c}x\\x_c\end{array}\right]\end{eqnarray*}
 is also pseudo strict bounded real.  This completes the proof of Theorem \ref{theorem20}. $\square$
\subsection{Proof of Theorem \ref{theorem200}:}
Consider the system described by the state equations
 \begin{equation} \label{sys20dual}
\begin{array}{ccl} \dot{\tilde{x}} & = &\tilde{A} \tilde{x}+\tilde{B}_2 \tilde{u}+\tilde{B}_1 \tilde{w},  \\
            \tilde{z}&=& \tilde{C}_1\tilde{x}+\tilde{D}_{12}\tilde{u},\\
            \tilde{y} &=& \tilde{C}_2 \tilde{x}+\tilde{D}_{21}\tilde{w},\\
            \end{array}
\end{equation}
where \begin{eqnarray}\label{matrices}  
\tilde{A} &=& A^T,~ \tilde{B}_1 = C_1^T, ~\tilde{B}_2 =C_2^T,
~ \tilde{C}_1 = B_1^T, ~
  \tilde{D}_{12} = D_{21}^T, \nonumber \\
\tilde{C}_2  &=& B_2^T, ~\tilde{D}_{21} = D_{12}^T,
\end{eqnarray}
Let \begin{equation}\label{XandY} \tilde{E}_1=\tilde{D}_{12}^T\tilde{D}_{12}=E_2,
~\tilde{E}_2=\tilde{D}_{21}^T\tilde{D}_{21}=E_1, ~\tilde{X}=Y, ~\tilde{Y}=X.\end{equation}

Substituting the matrices in (\ref{matrices}) and (\ref{XandY}) into Conditions (i), (ii), (iii) of
the theorem gives that the system (\ref{sys20dual}) satisfies the following conditions of Theorem
\ref{theorem20}:

\begin{description}
\item[(i')] The Riccati, as shown below,  has a pseudo-positive definite stabilizing solution
\begin{eqnarray}
\label{AREt1} 
\lefteqn{(\tilde{A}-\tilde{B}_2 \tilde{E}_1^{-1} \tilde{D}_{12}^T \tilde{C}_1)^T \tilde{X} +
\tilde{X} (\tilde{A}-\tilde{B}_2 \tilde{E}_1^{-1} \tilde{D}_{12}^T\tilde{C}_1)}\nonumber \\
&&+\tilde{X}(\tilde{B}_1 \tilde{B}_1^T-\tilde{B}_2\tilde{E}_1^{-1}\tilde{B}_2^T)\tilde{X} \nonumber\\
&&+ \tilde{C}_1^T(I-\tilde{D}_{12}\tilde{E}_1^{-1}\tilde{D}_{12}^T)\tilde{C}_1=0. \hspace{2.5cm}
\end{eqnarray}

\item[(ii')]
The following Riccati equation  has a positive definite stabilizing solution
\begin{eqnarray}
\label{AREt2}
\lefteqn{(\tilde{A}-\tilde{B}_1\tilde{D}_{21}^T\tilde{E}_2^{-1}\tilde{C}_2)\tilde{Y}
+\tilde{Y}(\tilde{A}-\tilde{B}_1\tilde{D}_{21}^T\tilde{E}_2^{-1}\tilde{C}_2)^T}\nonumber \\
&&+\tilde{Y}(\tilde{C}_1^T\tilde{C}_1-\tilde{C}_2^T\tilde{E}_2^{
-1} \tilde{C}_2)\tilde{Y}\nonumber\\
&&+ \tilde{B}_1
(I-\tilde{D}_{21}^T\tilde{E}_2^{-1}\tilde{D}_{21})\tilde{B}^T_1=0. \hspace{2.5cm}
\end{eqnarray}

\item[(iii')] The matrix $\tilde{X}\tilde{Y}$ has a spectral radius strictly less than one, $\rho(\tilde{X}\tilde{Y})<1$.
\end{description}
Using Theorem \ref{theorem20}, it follows that there exists a dynamic output feedback compensator
of the form (\ref{Ch6controller}) such that the closed-loop system consisting of the system
(\ref{sys20dual}) and this compensator is pseudo strict bounded real. 
The parameters of this compensator are as follows:
\begin{eqnarray}\label{controllerParametersTilde}\tilde{A}_c &=& \tilde{A}+\tilde{B}_2\tilde{C}_c-\tilde{B}_c\tilde{C}_2+(\tilde{B}_1-\tilde{B}_c\tilde{D}_{21})\tilde{B}_1^T\tilde{X},\nonumber \\
\tilde{B}_c &=& (I-\tilde{Y}\tilde{X})^{-1}(\tilde{Y}\tilde{C}_2^T+\tilde{B}_1\tilde{D}_{21}^T)\tilde{E}_2^{-1},\nonumber\\
\tilde{C}_c &=&
-\tilde{E}_1^{-1}(\tilde{B}_2^T\tilde{X}+\tilde{D}_{12}^T\tilde{C}_1).
\end{eqnarray}

Substituting the matrix in (\ref{matrices}) and (\ref{XandY}) into
(\ref{controllerParametersTilde}), the transfer function of this closed-loop system becomes $
\tilde {G}(s)=\left[\begin{array}{cc} B_1^T &
D_{21}^T B_c^T\end{array}\right]\left(s I-\left[\begin{array}{cc}A^T & C_2^T B_c^T \\
C_c^T B_2^T & A^T_c \end{array}\right] \right)^{-1} \left[\begin{array}{c} C_1^T \\
C_c^T D_{12}^T\end{array} \right]$.

Consider the system (\ref{Ch6sys3}) with compensator (\ref{Ch6controller}) whose parameters are
determined by (\ref{controllerParameters}). It is readily seen that the transfer function of this
closed-loop system $\check{G}(s)$ satisfies $\check{G}(s)=\tilde {G}^T(s)$.
 Therefore, from the fact that the system
(\ref{sys20dual}), (\ref{matrices}), (\ref{XandY}), (\ref{controllerParametersTilde}) is pseudo
strict bounded real, it follows that $\max\limits_{\omega}\sigma_{max}
[\check{G}^T(-j\omega)\check{G}(j\omega)]<1$. Also, $ \left[\begin{array}{cc}\tilde{ A} & \tilde{B}_2 \tilde{C}_c \\
\tilde{B}_c \tilde{C}_2 & \tilde{A}_c \end{array}\right]^T=\left[\begin{array}{cc} A & B_2 C_c \\
B_c C_2 & A_c \end{array}\right]$ and is pseudo-Hurwitz.   Hence, the closed-loop system
(\ref{Ch6sys3}), (\ref{Ch6controller}), (\ref{controllerParameters}) is pseudo strict bounded
real.$\square$
 \subsection{Proof of Theorem \ref{Ch6.OF.Theorem.B}:} We first prove that the closed-loop system
\begin{eqnarray}\label{Ch6Aug.OF.B}
\left[ \begin{array}{c}
       \dot{x}_c \\
       \dot{x}
      \end{array}
\right] & = & \left[ \begin{array}{cc}A_c & B_c C_2 \\ B_2 C_c  & A
\end{array}\right]\left[ \begin{array}{c}
       x_c \\x
      \end{array}
\right] +\left[ \begin{array}{c} B_c D_{21}
\\B_1\end{array} \right]w,\nonumber\\
        z & = & \left[ \begin{array}{cc} D_{12} C_c & C_1\end{array}\right]\left[ \begin{array}{c}
       x_c\\ x
      \end{array}
\right], \end{eqnarray}obtained by substituting the controller (\ref{Ch6controller}),
(\ref{Ch6AREparameter.B}) into the system (\ref{Ch6sys3}), is pendulum-like.  Let $
\bar{d}=[\begin{array}{cc}
                                            0_{1\times n} &
                                             e_n^T T^T
                                          \end{array}
]^T$.  Note the identity 
\begin{eqnarray*}
\lefteqn{\left[\begin{array}{cc}
                                               I & 0 \\
                                               0 & T
                                             \end{array}
\right]^{-1}\left[\begin{array}{cc}A_c & B_c C_2 \\ B_2 C_c  & A
\end{array}\right]\bar{d}}\nonumber \\
&=&\left[\begin{array}{ccc}
                                  A_c & B_c C_{2a} & 0 \\
                                  B_{2a}C_c & \tilde{A}_1 & 0 \\
                                  B_{2b}C_c & \tilde{A}_2 & 0
                                \end{array}
\right]\left[\begin{array}{c}
               0_{(2n-1)\times 1} \\
               1
             \end{array}
\right]=0.
\end{eqnarray*}
 Since $\left[ \begin{array}{cc}
                                               I & 0 \\
                                               0 & T
                                             \end{array}
\right]$ is non-singular, it follows that $\left[ \begin{array}{cc}A_c & B_c C_2 \\
B_2 C_c & A
\end{array}\right]\bar{d}=0$. Using this fact and Assumption \ref{Ch6.Assum.Unobserv}, it follows from
Lemma \ref{rationalPendulum} that the resulting closed-loop system (\ref{Ch6Aug.OF.B}) is
pendulum-like system with respect to the set $\Pi(\tau_0\bar{p}\bar{d})$.

From the output feedback pseudo $H_{\infty}$ control theory in  Section III,
Conditions I, II, III of the theorem imply that the matrix $\left[ \begin{array}{cc} \lambda I+A & B_2 C_c \\
B_c C_2 & \lambda I+ A_c \end{array}\right]$ is pseudo-Hurwitz and the frequency-domain condition
$\max\limits_{\omega}\sigma_{max} [\bar{G}^T(-j\omega)\bar{G}(j\omega)]<1$ holds, where
$\bar{G}(\cdot)$ is defined as $\bar{G}(s)\dfn M_{\tau}^{\frac{1}{2}}G_c(s)M_{\tau}^{-\frac{1}{2}}$
and here $G_c(s)\dfn\left[ \begin{array}{cc}
 C_1 &
D_{12} C_c\end{array}\right]\left(s-\left[ \begin{array}{cc} \lambda I+A & B_2 C_c \\
B_c C_2 &\lambda I+ A_c \end{array}\right] \right)^{-1} \left[ \begin{array}{c} B_1
\\B_c D_{21}\end{array} \right].$
  Then, it follows that
$G_c^T(-j\omega)M_{\tau}G_c(j\omega) <M_{\mu}^{-1}M_{\tau}M_{\mu}^{-1}$ for all
$\omega\in\mathcal{R}$. Now, all the conditions of Lemma \ref{LagrangeStability} are satisfied and
hence the closed-loop nonlinear system (\ref{Ch6Aug.OF.B}), (\ref{wi}), (\ref{nonlinearPeriod}),
(\ref{Ch6sector_1}) is Lagrange stable.  $\square$

\subsection{Proof of Theorem \ref{Ch6.OF.Theorem.A}: }
We first prove that the closed-loop system (\ref{Ch6Aug.OF.B}), obtained by applying the
compensator (\ref{Ch6controller}), (\ref{Ch5controllerP1}) to the system (\ref{Ch6sys3}), is a
pendulum-like system.

Since $\left[ \begin{array}{cc}I & 0\\ 0 & \bar{T}\end{array}\right]^{-1}\left[ \begin{array}{cc} A_c & B_c C_2\\
 B_2 C_c &  A\end{array}\right]\bar{d}
 = \left[ \begin{array}{cc}\left[ \begin{array}{cc}A_c & B_c\tilde{C}_{2a}\\ \tilde{B}_{2a} &
\tilde{A}_1\end{array}\right] &
\left[ \begin{array}{c}B_c\tilde{C}_{2b}\\\tilde{A}_2\end{array}\right]\\
0_{1\times (n-1)} & 0
\end{array} \right]\left[ \begin{array}{c} \bar{d}_0 \\1
\end{array} \right]=0$ ~~~
 and $\left[ \begin{array}{cc}I & 0\\
0 & \bar{T}\end{array}\right]$ is a non-singular matrix, it follows that $\left[ \begin{array}{cc}A_c & B_c C_2\\
 B_2 C_c &  A \end{array}\right]\bar{d}=0$.   Using this fact and Condition IV of the theorem, it
follows from Lemma \ref{rationalPendulum} that the augmented closed-loop system
(\ref{Ch6Aug.OF.B}), (\ref{wi}), (\ref{nonlinearPeriod}), (\ref{Ch6sector_1}) is a pendulum-like
system with respect to $\Pi(\bar{p}\tau_0\bar{d})$.

Using the output feedback pseudo $H_{\infty}$ control theory given in Section III, it follows from
Conditions I, II and III of the theorem that the closed-loop system (\ref{Ch6Aug.OF.B}) is pseudo
strict bounded real. In a similar way to the proof of Theorem \ref{Ch6.OF.Theorem.B}, we have
$G_c^T(-j\omega-\lambda)M_{\tau}G_c(j\omega-\lambda) <M_{\mu}^{-1}M_{\tau}M_{\mu}^{-1}$. Now, using
Lemma \ref{LagrangeStability},  it follows that the closed-loop system (\ref{Ch6Aug.OF.B}),
(\ref{wi}), (\ref{nonlinearPeriod}), (\ref{Ch6sector_1}) is Lagrange stable. $\square$

\subsection{Proof of Theorem \ref{Ch6RationalityTest}}  The stabilizing solutions to the Riccati equations (\ref{Ch6ARE1}) and
(\ref{Ch6ARE2}) are functions of the vector of constants $\bar{\tau}$. To highlight this, we use
the notation ${X}(\bar{\tau})$ and ${Y}(\bar{\tau})$.
 In the proof of Theorem \ref{Ch6RationalityTest}, we use the following lemma:
\begin{lemma}\label{Ch6claimB} The nonsingular stabilizing solutions $X(\bar{\tau})$ and
 $Y(\bar{\tau})$ to Riccati equations
(\ref{Ch6ARE1}) and (\ref{Ch6ARE2}) are real analytic functions on the set $\mathbb{T}$.
\end{lemma}  \emph{Proof:} As $X(\bar{\tau})$ is nonsingular, we can rewrite the Riccati equation
(\ref{Ch6ARE1}) as
\begin{eqnarray}
\label{Ch6ARE1stab}\lefteqn{ (\lambda I+A-B_2 \bar{E}_1^{-1} D_{12}^T M_{\tau}C_1)
 +B_1M_{\mu}M_{\tau}^{-1}M_{\mu}B_1^TX(\bar{\tau})}\nonumber\\
&& -B_2\bar{E}_1^{-1}B_2^TX(\bar{\tau}) \nonumber \\
&=&-X^{-1}(\bar{\tau})\left[\begin{array}{l}(\lambda I+A-B_2 \bar{E}_1^{-1} D_{12}^T M_{\tau} C_1)^T
+\\
C_1^T(M_{\tau}-M_{\tau}D_{12}\bar{E}_1^{-1}D_{12}^T
M_{\tau})C_1X^{-1}(\bar{\tau})\end{array}\right]\nonumber \\
&&\hspace{1cm}\times X(\bar{\tau}).
\end{eqnarray}
As $X(\bar{\tau})$ is a pseudo-positive definite stabilizing solution to the Riccati equation (\ref{Ch6ARE1}), it follows that the matrix 
$ -(\lambda I+A-B_2 \bar{E}_1^{-1} D_{12}^T M_{\tau}C_1)^T    -
C_1^T(M_{\tau}-M_{\tau}D_{12}\bar{E}_1^{-1}D_{12}^T M_{\tau})C_1X^{-1}(\bar{\tau})$ is Hurwitz  and
hence the pair $(-(\lambda I+A-B_2 \bar{E}_1^{-1} D_{12}^T M_{\tau} C_1)^T,
-C_1^T(M_{\tau}-M_{\tau}D_{12}\bar{E}_1^{-1}D_{12}^T M_{\tau})C_1)$ is stabilizable.

The Riccati equation (\ref{Ch6ARE1}) can be written as
\begin{eqnarray}
\label{Ch6ARE1inv}\lefteqn{  X^{-1}(\bar{\tau})(\lambda I+A-B_2 \bar{E}_1^{-1} D_{12}^T M_{\tau}
C_1)^T}\nonumber \\
&&+ (\lambda I+A-B_2 \bar{E}_1^{-1} D_{12}^T M_{\tau}C_1)X^{-1}(\bar{\tau})
 \nonumber\\
&&+(B_1M_{\mu}M_{\tau}^{-1}M_{\mu}B_1^T-B_2\bar{E}_1^{-1}B_2^T)\nonumber \\
&&+
X^{-1}(\bar{\tau})C_1^T(M_{\tau}-M_{\tau}D_{12}\bar{E}_1^{-1}D_{12}^T
M_{\tau})C_1X^{-1}(\bar{\tau})=0. \nonumber \\
\end{eqnarray}
 Substituting the matrices $-(\lambda I+A-B_2 \bar{E}_1^{-1} D_{12}^T M_{\tau}
C_1)^T$,~ $C_1^T(M_{\tau}-M_{\tau}D_{12}\bar{E}_1^{-1}D_{12}^T M_{\tau})C_1$ ~and ~
$-B_1M_{\mu}M_{\tau}^{-1}M_{\mu}B_1^T + B_2\bar{E}_1^{-1}B_2^T$ into $A$, $R$ and $Q$ of Theorem 2
in \cite{Wimmer1985}, respectively, it follows that $X^{-1}(\bar{\tau})$ is the maximal solution
for all solutions of the Riccati equation (\ref{Ch6ARE1inv}). Since
$C_1^T(M_{\tau}-M_{\tau}D_{12}\bar{E}_1^{-1}D_{12}^T M_{\tau})C_1\ge 0$ and
$-B_1M_{\mu}M_{\tau}^{-1}M_{\mu}B_1^T+B_2\bar{E}_1^{-1}B_2^T$ is Hermitian,
Theorem 4.1 in \cite{Ran1988} is applicable. Using  
Theorem 4.1 in \cite{Ran1988} by substituting $-(\lambda I+A-B_2 \bar{E}_1^{-1} D_{12}^T M_{\tau}
C_1)^T$, $C_1^T(M_{\tau}-M_{\tau}D_{12}\bar{E}_1^{-1}D_{12}^T M_{\tau})C_1$ and
$-B_1M_{\mu}M_{\tau}^{-1}M_{\mu}B_1^T+B_2\bar{E}_1^{-1}B_2^T$ into $A$, $R$ and $Q$, respectively,
gives $X^{-1}(\bar{\tau})$ is a real analytic function of
$\bar{\tau}\in\mathbb{T}$. This further implies that $X(\bar{\tau})$ is a real analytic function of $\bar{\tau}\in\mathbb{T}$. 
Similarly, we can verify that $Y(\bar{\tau})$ is also a real analytic function of
$\bar{\tau}\in\mathbb{T}$.$\square$

\emph{Proof of Theorem \ref{Ch6RationalityTest}:}
Let $\epsilon>0$ be chosen to be sufficiently small so that the set
$\mathbb{B}(\tilde{\tau},\epsilon)=\left\{\bar{\tau}\in\mathcal{R}^m_+: \|\bar{
\tau}-\tilde{\tau}\|< \epsilon \right\}
 \subset\left\{\tilde{\tau} \in\mathbb{F}: \textrm{Condition}\right.$ $\left.\textrm{I, II and III of Theorem \ref{Ch6.OF.Theorem.A} holds} \right\}.$ The existence of such an
 $\epsilon>0$ follows from Lemma \ref{Ch6claimB}.

 Since $ X(\bar{\tau})$ and $ Y(\bar{\tau})$ are analytic function on the set $\mathbb{T}$, it straightforward to verify that
$f(\tilde{\tau})$ is an analytic function on the set $\mathbb{F}$. Since $\Delta^{-1}$ is a
diagonal positive definite matrix, it follows that Condition II of the theorem implies that $\det
J(\tilde{\tau})\neq 0$. Let ${c}=f\left(\tilde{\tau}\right)$. It follows from the Inverse Function
Theorem (e.g., see Theorem 7.8 in \cite{Haaser1971}) that there is an open ball
$\mathbb{B}(c,\iota)$ and a unique continuously differentiable function $g$ from
$\mathbb{B}(c,\iota)$ into $\mathbb{B}(\tilde{\tau},\epsilon)$ such that $ \tilde{\tau}=g(c)$
 and $f(g(\bar{c}))=\bar{c}$ for all $\bar{c}\in\mathbb{B}(c,\iota)$.

 Since the set of rational vectors $\mathcal{Q}^m$ is dense in $\mathcal{R}^m$, we can choose  $\check{c}\in\mathbb{B}(c,\iota)$
 such that all the elements of $\check{c}$ are rational and non-zero. Also, it follows from the above
 discussion that there exists a point $\check{\tau} \in \mathbb{B}(\tilde{\tau},\epsilon)$ such that
 $f(\check{\tau})=\check{c}$ where
 $\check{\tau}=g(\check{c})$. Therefore, Condition IV of Theorem
 \ref{Ch6.OF.Theorem.A} is satisfied.

It follows from the definition of $\mathbb{B}(\tilde{\tau},\epsilon)$ that $\check{\tau}$ satisfies
Conditions I, II and III of Theorem \ref{Ch6.OF.Theorem.A}. Hence,   Theorem \ref{Ch6.OF.Theorem.A}
implies that the corresponding closed-loop system is pendulum-like and Lagrange stable. $\square$

\subsection{Proof of Theorem \ref{Ch6StateFDsufficient}:}
 Substituting the controller law (\ref{Ch6StateFeedback}) into the system (\ref{Ch6sys3a}), (\ref{Ch6sys3b}) gives
the closed-loop system
\begin{eqnarray}
\label{Ch6sys.SF.Clsd}
   \dot{x} &=& \left(A-B_2\bar{E}_1^{-1}D_{12}^T M_{\tau} C_1
 -B_2\bar{E}_1^{-1}B_2^TX\right)x+ B_1\xi, \nonumber \\
   z  &=& \left(\left(I- D_{12} \bar{E}_1^{-1}D_{12}^T M_{\tau}\right) C_1 -
D_{12} \bar{E}_1^{-1}B_2^TX\right) x.
\end{eqnarray}
Since $T^{-1}(A-B_2\bar{E}_1^{-1}D_{12}^T M_{\tau}C_1 -B_2\bar{E}_1^{-1}B_2^TX)\bar{d}
=(\tilde{A}- \tilde{B}_2\bar{E}_1^{-1}D_{12}^T M_{\tau}\tilde{C}_1
-\tilde{B}_2\bar{E}_1^{-1}\tilde{B}_2^T\bar{X})\bar{d}=0$,   
it follows that $(A - B_2\bar{E}_1^{-1}D_{12}^T M_{\tau}C_1 -B_2\bar{E}_1^{-1}B_2^TX)\bar{d}=0$.
Using this fact and condition II, it follows from Lemma \ref{rationalPendulum} that the closed-loop
system \eqref{Ch6sys.SF.Clsd} is a pendulum-like system with respect to
$\Pi(\bar{p}\tau_0\bar{d})$.

Using the fact that the Riccati equation (\ref{Ch6ARE1}) has a pseudo-positive definite solution
and Condition I holds,  Theorem \ref{synthesis0} implies
that the closed-loop system (\ref{Ch6sys.SF.Clsd}) is pseudo strict bounded real. 
Then, using Lemma \ref{LagrangeStability}, it follows that the closed-loop system
(\ref{Ch6sys.SF.Clsd}), (\ref{Ch6sys3b}), (\ref{wi}), (\ref{nonlinearPeriod}), (\ref{Ch6sector_1})
 is Lagrange stable.  $\square$
\begin{biography}[{\includegraphics[width=1in,height=1.25in,clip,keepaspectratio]{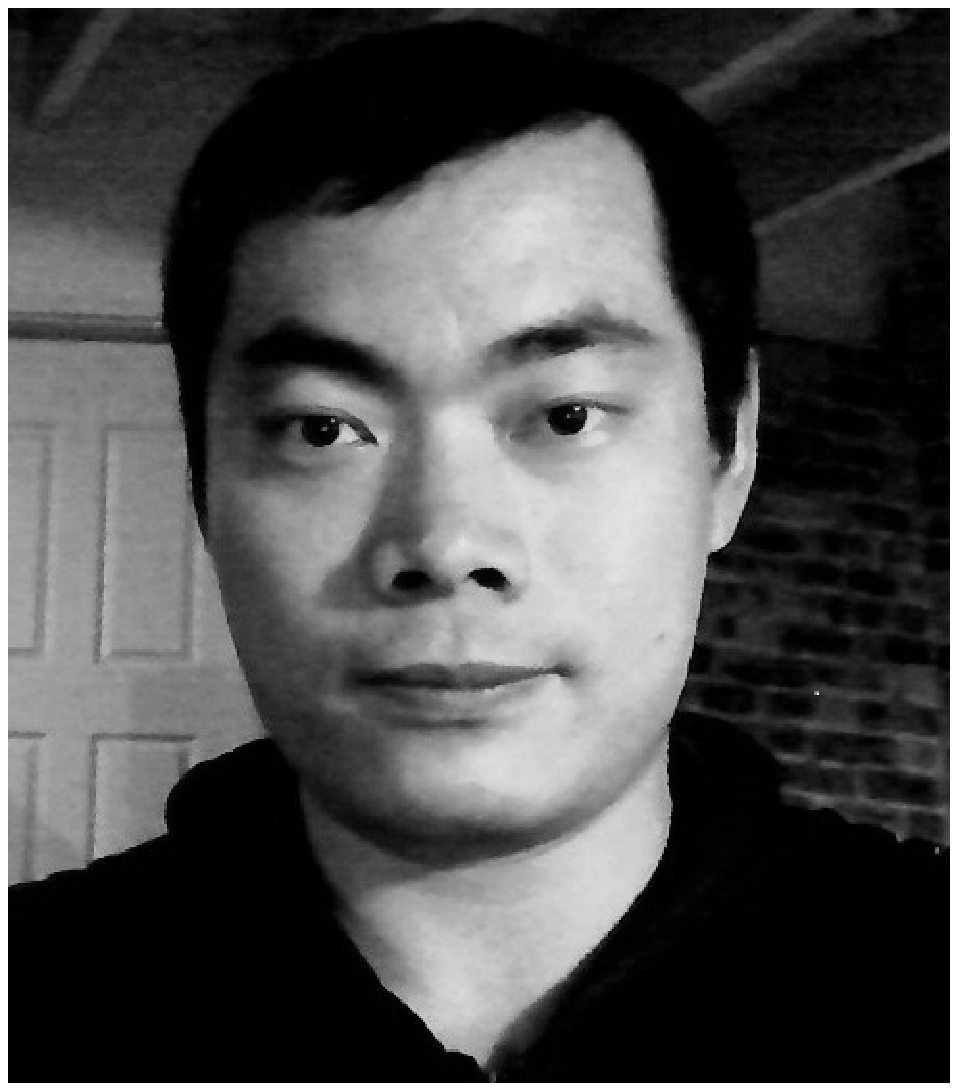}}]{Hua Ouyang}
Hua Ouyang received his B. Eng in Engineering Mechanics from Hunan University, China and M.S. degree in Dynamics and Control from Peking University, China in 2000 and 2003, respectively.  He obtained his first Ph.D. degree in Industrial Electronics from the University of Glamorgan, UK and his second Ph.D. degree in Control Theory and Applications from the University of New South Wales at the Australian Defence Force Academy in 2007 and 2011, respectively. Now, he is a research fellow in the School of Chemical Engineering, University of New South Wales, Australia.  His research interests include robust control and filtering, networked control systems and nonlinear systems (feedback linearization, pendulum-like systems and flow control).  
\end{biography}

\begin{biography}[{\includegraphics[width=1in,height=1.25in,clip,keepaspectratio]{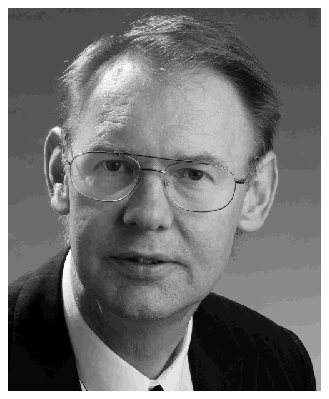}}]{Ian Petersen}
 was born in Victoria, Australia. He received a
  Ph.D in Electrical Engineering in 1984 from the University of
 Rochester. From 1983 to 1985 he was a Postdoctoral Fellow at the
 Australian National University. In 1985 he joined the  University of
 New South Wales at the 
Australian Defence Force Academy where he is currently Scientia 
Professor and an Australian Research Council Federation Fellow in the
 School of  Information Technology and Electrical Engineering. He has
 served as an Associate Editor for the IEEE Transactions on Automatic
 Control, Systems and Control Letters, Automatica, and SIAM Journal on
 Control and Optimization. Currently he is 
an Editor for Automatica. He is a fellow of the IEEE and the Australian Academy 
of Sciences.  His main
 research interests are in robust control theory, quantum control
 theory and stochastic control theory. 
\end{biography}
\begin{biography}[{\includegraphics[width=1in,height=1.25in,clip,keepaspectratio]{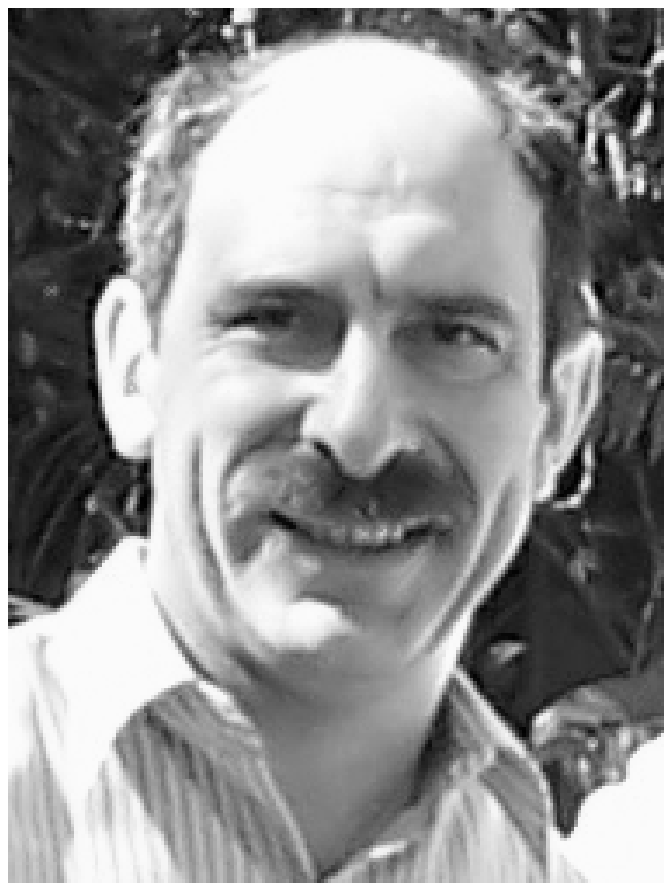}}]{Valery Ugrinovskii} (M'97-SM'02) received the undergarduate degree in
applied mathematics and the PhD degree in physics and mathematics from the
State University of Nizhny Novgorod, Russia, in 1982 and 1990,
respectively.

He is currently an Associate Professor in the School of Engineering and
Information Technology, at the University of New South Wales at the
Australian Defence Force Academy, in Canberra. From 1982 to 1995, he held
research positions with the Radiophysical Research Institute, Nizhny
Novgorod. From 1995 to 1996, he was a Postdoctoral Fellow at the University
of Haifa, Israel. In 2005, he held visiting appointments at the Australian
National University. He is the coauthor of the research monograph
\emph{Robust Control Design using $H^\infty$ Methods}, Springer, London,
2000, with Ian R. Petersen and Andrey V. Savkin. His current research
interests include decentralized and distributed control, stochastic control
and filtering theory, robust control, and switching control.

Dr. Ugrinovskii serves as an Associate Editor of Automatica.
\end{biography}
\end{document}